\documentclass[useAMS,usenatbib]{mn2e}
\usepackage[dvips]{graphicx}
\usepackage[english]{babel}
\usepackage{amsmath}
\usepackage{amssymb}
\usepackage{textcomp}
\usepackage{natbib}

\title[Soft X-ray and UV emission from halo gas]{Soft X-ray and ultra-violet metal-line emission from the gas around galaxies}
\author[F. van de Voort and J. Schaye]{Freeke van de Voort$^{1}$\thanks{E-mail: fvdvoort@strw.leidenuniv.nl} and Joop Schaye$^{1}$ \\
$^{1}$Leiden Observatory, Leiden University, Postbus 9513, 2300 RA, Leiden, The Netherlands}

\begin{document}

\date{Accepted not yet. Received \today; in original form July 25, 2012}

\pagerange{\pageref{firstpage}--\pageref{lastpage}} \pubyear{2013}

\maketitle

\label{firstpage}

\begin{abstract}
A large fraction of the gas in galactic haloes has temperatures between $10^{4.5}$ and $10^7$~K. At
these temperatures, cooling is dominated by metal-line emission if the metallicity $Z\gtrsim0.1$~$Z_\odot$ and several lines may be detectable with current and upcoming instruments. We explore this possibility using several large cosmological, hydrodynamical simulations from the OverWhelmingly Large Simulations project. We stack surface brightness maps centred on galaxies to calculate the expected mean surface brightness profiles for different halo masses. We only consider emission from intergalactic gas with densities $n_{\rm H} < 0.1$~cm$^{-3}$. Results are shown for soft X-ray lines at $z=0.125$, ultra-violet (UV) lines and hydrogen Balmer-$\alpha$ (H$\alpha$) at $z=0.25$, and rest-frame UV lines at $z=3$. Assuming a detection limit of
$10^{-1}$~photon~s$^{-1}$\,cm$^{-2}$\,sr$^{-1}$, proposed X-ray telescopes can detect O\,\textsc{viii} emission from $z=0.125$ out to 80 per cent of the virial radius ($R_\mathrm{vir}$) of groups and clusters and out to 0.4$R_\mathrm{vir}$ for haloes with masses 10$^{12-13}$~M$_\odot$. Emission lines from C\,\textsc{vi}, N\,\textsc{vii}, O\,\textsc{vii}, and Ne~\textsc{x} can be detected out to smaller radii, $0.1-0.5R_\mathrm{vir}$. With a detection limit of $10^{-20}$~erg~s$^{-1}$\,cm$^{-2}$\,arcsec$^{-2}$, future UV telescopes can detect C\,\textsc{iii} emission out to 0.2--0.6$R_\mathrm{vir}$ at $z=0.25$, depending on halo mass. C\,\textsc{iv}, O\,\textsc{vi}, Si\,\textsc{iii}, and Si\,\textsc{iv} can be seen out to $10-20$ per cent of the virial radius in haloes more massive than $10^{12}$~M$_\odot$. Optical H$\alpha$ emission is comparable in strength to C\,\textsc{iii} emission and could be observed out to 0.3--0.6$R_\mathrm{vir}$ at $z=0.25$ with upcoming optical instruments. At $z=3$ it may be possible to observe C\,\textsc{iii} out to $0.2-0.3R_\mathrm{vir}$ and other rest-frame UV lines out to $\sim0.1R_\mathrm{vir}$ in haloes larger than $10^{11}$~M$_\odot$ with the same optical instruments. 
Metal-line emission is typically biased towards high density and metallicity and towards the temperature at which the emissivity curve of the corresponding metal line peaks. This bias varies with radius, halo mass, and redshift. The bias is similar for the different soft X-ray lines considered, whereas it varies strongly between different UV lines.
Active galactic nucleus (AGN) feedback can change the inner surface brightness profiles significantly, but it generally does not change the radius out to which the emission can be observed. Metal-line emission is a promising probe of the warm and hot, enriched gas around galaxies and provides a unique window into the interactions between galaxies and their gaseous haloes.
\end{abstract}


\section{Introduction}

Haloes grow by accreting gas from their surroundings, the intergalactic medium (IGM), which is the main reservoir of baryons \citep[see e.g.][]{Shull2012}. Galaxies grow by accreting gas from their haloes, from which they can form stars. Some of the gas is returned to the circumgalactic medium (CGM) in galactic winds driven by supernovae or active galactic nuclei (AGN). Metals produced in stars are also blown out of the galaxy and enrich the CGM \citep[see e.g.][]{Stinson2012}. Hence, to understand the evolution of galaxies, one needs to study the evolution of the halo gas.

Ultra-violet (UV) and X-ray absorption line studies have revealed both the cold, neutral gas and the warm-hot intergalactic medium around galaxies \citep[e.g.][]{Nicastro2005b, Tumlinson2011, Prochaska2011, Williams2013}. Unfortunately, this type of observation can only provide information about the gas along the line of sight to the background quasar, so there is no information about the transverse extent of the absorbing gas cloud. Another limitation is that we can only probe gas in case there is a bright background object. Therefore, although absorption line studies are excellent for studying regions of moderate overdensity in the IGM, the statistics are poorer for relatively dense halo gas, particularly around massive haloes.

The gas emissivity scales with the square of the density. Emission is thus dominated by high-density regions and is thus a better tracer of the CGM than of the general IGM. Emission line studies thus complement absorption line studies. They have the added advantage of providing a two-dimensional image, in addition to the third dimension provided by the redshift of the emission line, allowing us to study the three-dimensional spatial distribution.

H~\textsc{i} Lyman-$\alpha$ (Ly$\alpha$) emission originates from regions with temperature $T\approx10^4$~K. It therefore provides an excellent route to studying cold gas in the CGM. Diffuse Ly$\alpha$ emission has been detected (statistically) around high-redshift galaxies at $z\sim3$ \citep[e.g.][]{Steidel2000, Steidel2011, Matsuda2004, Bower2004}. However, depending on halo mass, a large fraction of the halo gas is expected to heat to temperatures above $10^{4.5}$~K, either through photo-ionization, accretion shocks, or through shocks caused by galactic winds \citep[e.g.][]{VoortSchaye2012}. UV metal-line emission enables us to probe gas with $T=10^{4.5-5.5}$~K, which tends to be more diffuse than colder Ly$\alpha$ emitting gas and thus a better probe of the average gas properties around galaxies \citep[e.g.][]{Bertone2012}. Soft X-ray metal-line emission traces even hotter gas with $T=10^{6-7}$~K, which is the dominant temperature of the halo gas around high-mass galaxies and in galaxy groups. At $T\sim10^{4.5-7}$~K the emission is dominated by metal lines for $Z\gtrsim0.1$~Z$_\odot$ \citep[e.g][]{Wiersma2009a} rather than by hydrogen lines or continuum emission as is the case for lower and higher temperatures, respectively.

Dilute halo gas that has been shock-heated to the virial temperature is routinely detected in X-ray observations of the centres of clusters and groups of galaxies, and it may even have been seen around individual galaxies \citep[e.g.][]{Crain2010a, Crain2010b, Anderson2011, Li2012}. Most of the detections are made in X-ray continuum emission and iron lines, see \citet{Bohringer2010} for a recent review, but many other X-ray lines have been identified. In the last year, a lot of progress has been made on observing X-ray emission close to the virial radii of clusters \citep{Simionescu2011, Akamatsu2011, Miller2012, Urban2011}, but see also \citet{Eckert2011}. Additionally, there are claims of detections of metal-line emission from the warm-hot intergalactic medium \citep[e.g.][]{Kaastra2003, Takei2007, Werner2008, Galeazzi2012}, whereas other observations set upper limits at lower values than these claimed detections \citep[e.g.][and references therein]{Mitsuishi2012}.

Observing metal-line emission from diffuse halo gas would yield a lot of information about the physical state of the gas, the distribution of metals, and thus the cycle of gas between haloes and galaxies. In general, however, emission from gas outside of galaxies is faint and thus difficult to detect. Many missions have been proposed to study the diffuse halo gas in emission. The next generation of spectrographs should be able to detect certain metal lines. Motivated by future instrumentation and recent proposals, recent studies have quantified the expected emission using cosmological, hydrodynamical simulations \citep{Furlanetto2004, Bertone2010a, Bertone2010b, Takei2011, Frank2012, Bertone2012, Roncarelli2012}. These studies are focussed mostly on quantifying the emission expected from the warm-hot intergalactic medium in a large section of the Universe or in mock datacubes. 

It is possible to increase the signal-to-noise by stacking many observations, centred on galaxies. In this way we cannot study the halo gas around a single object, but we can characterize the general properties of gas around a certain type of galaxy. In this work we will use cosmological, hydrodynamical simulations from the overwhelmingly large simulations (OWLS) project \citep{Schaye2010} to quantify the expected surface brightness (SB) for the brightest metal lines for a range of halo masses, in the (rest-frame) UV at high ($z=3$) and low ($z=0.25$) redshift and in soft X-ray at low ($z=0.125$) redshift. Our study complements previous work on metal-line emission from the IGM by focussing on the CGM and by predicting mean surface brightness profiles for different halo masses.

This paper is organized as follows. In Section~\ref{sec:method} we describe the simulations we used, how we identify haloes, and how we calculate the emission signal. The detection limits of soft X-ray lines at $z=0.125$ and UV lines at $z=0.25$ and $z=3$ with current and future instruments are described in Section~\ref{sec:detect}. The obtained surface brightness profiles are shown and discussed in Section~\ref{sec:SB}. Section~\ref{sec:prop} discusses how the gas temperature, density, and metallicity of gas traced by metal-line emission are biased. The effect of AGN feedback on our results is discussed in Section~\ref{sec:AGN}. We conclude in Section~\ref{sec:concl} and resolution tests are discussed in the appendix.

In all our calculations we assume the same cosmological parameter values as were used during our simulations (see Section~\ref{sec:sim}: $\Omega_\mathrm{m} = 1 - \Omega_\Lambda = 0.238$ and $h=0.73$).

Data products constructed from our simulations are available upon request.

\section{Method} \label{sec:method}

\subsection{Simulations} \label{sec:sim}

We use a modified version of \textsc{gadget-3} \citep[last described in][]{Springel2005}, a smoothed particle hydrodynamics (SPH) code that uses the entropy formulation of SPH \citep{Springel2002}, which conserves both energy and entropy where appropriate. This work is part of the OWLS project \citep{Schaye2010}, which consists of a large number of cosmological simulations with varying (subgrid) physics. Note that the OWLS models were also used in the studies of metal-line emission by \citet{Bertone2010a, Bertone2010b, Bertone2012}. Here we make use of the so-called `reference' model, although we will compare the soft X-ray results with a simulation that includes AGN feedback. The model is fully described in \citet{Schaye2010} and we will only summarize its main properties here.

The simulations assume a $\Lambda$CDM cosmology with parameters $\Omega_\mathrm{m} = 1 - \Omega_\Lambda = 0.238$, $\Omega_\mathrm{b} = 0.0418$, $h = 0.73$, $\sigma_8 = 0.74$, $n = 0.951$. These values are consistent\footnote{The only significant discrepancy is in $\sigma_8$, which is 8 per cent, or 2.3$\sigma$, lower than the value favoured by the WMAP 7-year data.} with the WMAP year~7 data \citep{Komatsu2011}. In the present work, we use the simulation output at redshifts 3 and 0.125.

A cubic volume with periodic boundary conditions is defined, within which the mass is distributed over $512^3$ dark matter and as many gas particles. The box sizes (i.e.\ the length of a side of the simulation volume) of the simulations used in this work are 25, 50, and 100~comoving~$h^{-1}$Mpc. The (initial) particle masses for baryons and dark matter are $2.1\times10^6(\frac{L_\mathrm{box}}{25\ h^{-1}\mathrm{Mpc}})^3$~M$_\odot$ and $1.0\times10^7(\frac{L_\mathrm{box}}{25\ h^{-1}\mathrm{Mpc}})^3$~M$_\odot$, respectively. 
The gravitational softening length is 1.95~$(\frac{L_\mathrm{box}}{25\ h^{-1}\mathrm{Mpc}})$~comoving~$h^{-1}$kpc, i.e.\ 1/25 of the mean dark matter particle separation, but is limited to a maximum value of 0.5~$(\frac{L_\mathrm{box}}{25\ h^{-1}\mathrm{Mpc}})$~proper~$h^{-1}$kpc, which is reached at $z=2.91$.

Star formation is modelled according to the recipe of \citet{Schaye2008}. A polytropic equation of state $P_\mathrm{tot}\propto\rho_\mathrm{gas}^{4/3}$ is imposed for densities exceeding $n_\mathrm{H}=0.1$~cm$^{-3}$, where $P_\mathrm{tot}$ is the total pressure and $\rho_\mathrm{gas}$ the density of the gas. Gas particles with proper densities $n_\mathrm{H}\ge0.1$~cm$^{-3}$ and temperatures $T\le10^5$~K are moved onto this equation of state and can be converted into star particles. The star formation rate (SFR) per unit mass depends on the gas pressure and reproduces the observed Kennicutt-Schmidt law \citep{Kennicutt1998} by construction. Particles with densities above $n_\mathrm{H}=0.1$~cm$^{-3}$ are not taken into account for the results in this work.

Feedback from star formation is implemented using the prescription of \citet{Vecchia2008}. About 40 per cent of the energy released by type II supernovae is injected locally in kinetic form, while the rest of the energy is assumed to be lost radiatively. The initial wind velocity is 600~km~s$^{-1}$ and the initial mass loading is two, meaning that, on average, a newly formed star particle kicks twice its own mass in neighbouring gas particles. 

The abundances of eleven elements released by massive stars and intermediate mass stars are followed as described in \citet{Wiersma2009b}. We assume the stellar initial mass function (IMF) of \citet{Chabrier2003}, ranging from 0.1 to 100~M$_\odot$. As described in \citet{Wiersma2009a}, radiative cooling and heating are computed element by element in the presence of the cosmic microwave background radiation and the \citet{Haardt2001} model for the UV/X-ray background from galaxies and quasars, assuming the gas to be optically thin and in (photo-)ionization equilibrium.

Using the suite of OWLS models, \citet{Wiersma2011} have shown that galactic winds driven by supernovae are essential for the enrichment of the IGM. Close to the centre of a halo, most gas has been enriched to $Z>0.1$~Z$_\odot$, but at the virial radius the scatter is very large \citep{VoortSchaye2012}. The simulations largely agree with observations of rest-frame UV absorption lines, both at low ($z\approx0.25$) and high ($z\approx3$) redshift. This has been shown for O~\textsc{vi} and H~\textsc{i} at low redshift by \citet{Tepper2011, Tepper2012} and for H~\textsc{i} at high redshift by \citet{Altay2011}. \citet{Tepper2011} do find a lack of high equivalent width  O~\textsc{vi} absorbers, which they suggest, following \citet{Oppenheimer2009}, can be attributed to unresolved turbulence on subgrid scales. OWLS has not yet been used to study X-ray absorption lines.

Finally, we also show soft X-ray results for a simulation that includes AGN feedback (model \emph{AGN}, Section~\ref{sec:AGN}). Black holes grow via mergers and gas accretion and inject 1.5 per cent of the rest-mass energy of the accreted gas into the surrounding matter in the form of heat. The model is a modified version of the one introduced by \citet{Springeletal2005} and is described and tested in \citet{Booth2009}, who also demonstrate that the simulation reproduces the observed mass density in black holes and the observed scaling relations between black hole mass and both central stellar velocity dispersion and stellar mass. \citet{McCarthy2010} have shown that model \emph{AGN} reproduces the observed stellar mass fractions, star formation rates, and stellar age distributions in galaxy groups, as well as the thermodynamic properties of the intragroup medium.

\subsection{Identifying haloes} \label{sec:halo}

The first step towards finding gravitationally bound structures, is to identify dark matter haloes. These can be found using a Friends-of-Friends (FoF) algorithm. If the separation between two dark matter particles is less than 20 per cent of the average separation (the linking length $b=0.2$), they are placed in the same group.
Baryonic particles are linked to a FoF halo if their nearest dark matter neighbour is in that halo. We then use \textsc{subfind} \citep{Springel2001} to find the most bound particle of a FoF halo, which serves as the halo centre and also corresponds to the location of the central galaxy. We compute the virial radius, $R_\mathrm{vir}$, within which the average density agrees with that predicted by the top-hat spherical collapse model in a $\Lambda$CDM cosmology \citep{Bryan1998}. The halo mass, $M_\mathrm{halo}$, is the total mass contained inside $R_\mathrm{vir}$. 

Table \ref{tab:sim} lists which simulations and which halo mass bins are used for various lines and redshifts. It also lists the median halo mass, median stellar mass, median star formation rate, and the number of haloes in the different mass bins.

\begin{table*}
\caption{\label{tab:sim} \small List of the simulation box sizes and halo mass ranges used for different emission lines at different redshifts. Included are also the median halo mass, median stellar mass, median star formation rate, and the number of haloes in each halo mass bin.}
\begin{tabular}[t]{lllrrrrr}
\hline
\hline \\[-3mm]
Redshift & Band & Simulation & Log$_{10}M_\mathrm{halo}$ range & Log$_{10}M_\mathrm{halo}$ & Log$_{10}M_\mathrm{star}$ & SFR & \# haloes \\
         &      &            & (M$_\odot$)                  & (M$_\odot$)            & (M$_\odot$)            & (M$_\odot$\,yr$^{-1}$) & \\
\hline \\[-4mm]
0.125   & X-ray & REF 100 $h^{-1}$Mpc & 12--13 & 12.3 & 10.8 & 8.95 & 2771 \\
        &       &                 & 13--14 & 13.2 & 11.8 & 45.71 & 301 \\
        &       &                 & 14--15 & 14.2 & 12.6  & 144.25  & 18 \\
\hline \\[-4mm]
0.125   & X-ray & AGN 100 $h^{-1}$Mpc & 12--13 & 12.3 & 10.3 & 0.44   & 2650 \\
        &       &                 & 13--14 & 13.2 & 11.1 & 2.09  & 242 \\
        &       &                 & 14--15 & 14.2 & 12.1 & 48.67  & 16 \\
\hline \\[-4mm]
0.25   & UV \& H$\alpha$ & REF 50 $h^{-1}$Mpc & 11--12 & 11.3 & 9.3 & 0.07  & 2722 \\
        &    &                & 12--13 & 12.3 & 10.9 & 11.55 &  371 \\
        &    &                & 13--14 & 13.2 & 11.7 & 45.81 & 39 \\
\hline \\[-4mm]
3.0   & UV & REF 25 $h^{-1}$Mpc & 10--11 & 10.3 & 8.2 & 0.11 & 1857 \\
      &    &                & 11--12 & 11.3 & 9.5 & 4.24  & 117 \\
      &    &                & 12--13 & 12.2 & 11.0 & 110.61 & 5 \\
\hline
\end{tabular}
\end{table*}

\subsection{Emission} \label{sec:em}

We calculate the emissivities of the gas following \citet{Bertone2010a}. We only summarize the method here and refer the reader to \citet{Bertone2010a} for details on the procedure. The names, wavelengths, and energies of the lines considered in this work are given in Table~\ref{tab:lines}.

We created emissivity tables as a function of temperature, density, and redshift with the photo-ionization package \textsc{cloudy} (version c07.02.02), which was last described in \citet{Ferland1998}. The gas is assumed to be optically thin and in ionization equilibrium in the presence of the cosmic microwave background and the \citet{Haardt2001} UV background. These assumptions were also made when calculating cooling rates during the simulation \citep{Wiersma2009a} and are thus fully consistent with the simulation. We further assumed solar abundances when creating the tables, but the results are scaled to the abundances predicted by the simulation. The tables are created for temperatures $10^2$~K$<T<10^{8.5}$~K with $\Delta$log$_{10}T=0.05$ bins and densities $10^{-8}$~cm$^{-3}<n_\mathrm{H}<10$~cm$^{-3}$ with $\Delta$log$_{10}n_\mathrm{H}=0.2$ bins.

We calculate the emission only for gas with $n_\mathrm{H}<0.1$~cm$^{-3}$, which is therefore by definition not star forming. Our simulations do not resolve the multiphase interstellar medium at $n_\mathrm{H}\ge0.1$~cm$^{-3}$. As we are interested in determining the emission from diffuse gas in haloes, ignoring this gas is consistent with our aims. Note, however, that because of this omission, the mean surface brightness profiles presented in this work may underestimate the true mean profiles.

The emissivity, $\epsilon$, for an emission line is derived from the tables as a function of log$_{10}T$ and log$_{10}n_\mathrm{H}$ through log-linear interpolation. A gas particle's luminosity is then
\begin{equation}
L = \epsilon(z, T, n_\mathrm{H})\dfrac{m_\mathrm{gas}}{\rho}\dfrac{X_y}{X_y^\odot}
\end{equation}
in erg\,s$^{-1}$, where $m_\mathrm{gas}$ is its mass, $\rho$ is its density, $X_y$ is the `SPH-smoothed' mass fraction of the element corresponding to the emission line, and $X_y^\odot$ is the solar mass fraction of the same element. We use SPH-smoothed abundances as described by \citet{Wiersma2009b}. This is consistent with the simulation, because these smoothed abundances were also used in the simulation to compute the cooling rates.
The flux is 
\begin{equation}
F = \dfrac{L}{4\pi d_\mathrm{L}^2}
\end{equation}
in erg\,s$^{-1}$\,cm$^{-2}$, or
\begin{equation}
F = \dfrac{L}{4\pi d_\mathrm{L}^2}\dfrac{\lambda}{h_\mathrm{p}c}(1+z)
\end{equation}
in photon~s$^{-1}$\,cm$^{-2}$, with $d_\mathrm{L}$ the luminosity distance, $h_\mathrm{p}$ the Planck constant, $c$ the speed of light, and $\lambda$ the rest-frame wavelength of the emission line.

To calculate the emission profiles, we project the flux using a flux-conserving SPH interpolation scheme onto a two-dimensional grid, centred on the halo centre. The surface brightness is calculated by dividing the flux by the solid angle subtended by a pixel either in arcsec$^2$ or in sr, 
\begin{equation}
S_\mathrm{B}=F/\Omega.
\end{equation}

We will investigate the detectability of the emission lines listed in Table~\ref{tab:lines}, originating from the CGM. We chose these lines, because they are the brightest in the outer haloes of galaxies, although in the centres other lines may dominate. We will quantify the expected surface brightness as a function of radius for a range of halo masses. Our predictions for detectability should be considered with care. We did not take into account the velocities of the gas and thus the width of the lines and we ignored the effect of the Earth's atmosphere. Intervening absorption was also ignored. A robust study of detectability should produce datacubes, including noise, and analyze them using the same pipeline as the observations. This is beyond the scope of the present work in which we show and discuss the theoretically expected surface brightness profiles.

To mimic the effect of UV emission from local sources, we repeated our analysis after increasing the UV background by a factor of ten. The results are unaffected, because the bright emission is coming from gas that is collisionally ionized. We thus conclude that the inclusion of radiation from local sources is unimportant.

O~\textsc{vii} is a line triplet at 0.574, 0.568, and 0.561 keV. The first line is a resonance line, the second one an intercombination line and the last one a forbidden line. Photons at the resonance energy see larger optical depths and are thus more likely to be scattered in random directions, giving rise to lower observed fluxes or even absorption. The intercombination line is significantly weaker ($0.5-0.6$~dex) than the other two. The resonance line is only slightly stronger than the forbidden line (by about 0.1~dex) without taking into account attenuation. Although the actual 0.574~keV surface brightness can in principle be much lower due to resonant scattering, we estimated the maximum optical depth for the O~\textsc{vii} resonance line in a halo to be very small. Out of the haloes we considered, only 20 have at least one pixel with optical depth $\tau>0.1$ (using Equation~44 in \citealt{Kaastra2008} with a velocity dispersion of 100~km~s$^{-1}$). The line is optically thin in the vast majority of cases, so we do not expect resonant scattering to influence our results. Another effect that could play a role is charge exchange, caused by collisions between hydrogen atoms and oxygen ions, in which case one would find enhanced emission for the forbidden O~\textsc{vii} line \citep[e.g.][]{Liu2011, Liu2012}. We only show the surface brightness of the forbidden line at 0.561~keV, but note that if the emission at all three wavelengths is added together or if charge exchange is important, then the signal could be more than twice as strong.

The UV doublets, C~\textsc{iv}, O~\textsc{vi}, and Si~\textsc{iv} have flux ratios of 2:1. We consider only the strongest line in this paper, but the results for the weaker lines thus follow directly.

\section{Expected detection limits} \label{sec:detect}

\begin{table}
\caption{\label{tab:lines} \small List of ion, rest-frame wavelength, and rest-frame energy for the soft X-ray, optical, and UV emission lines used in this work.}
\begin{tabular}[t]{lrr}
\hline
\hline \\[-3mm]
ion & $\lambda$ (\AA) & E (eV)\\
\hline \\[-4mm]
C \textsc{vi}    & 33.74 & 367  \\
N \textsc{vii}   & 24.78 & 500  \\
O \textsc{vii}   & 22.10 & 561 \\
O \textsc{viii}  & 18.97 & 654  \\
Ne \textsc{x}    & 12.14 & 1021  \\
\hline \\[-4mm]
H \textsc{i} (H$\alpha$)   & 6563 &  1.89  \\
\hline
C \textsc{iii}   & 977  &  12.69   \\
C \textsc{iv}   & 1548 &  8.01 \\
O \textsc{vi}   & 1032 &  12.01\\
Si \textsc{iii}  & 1207 &  10.27  \\
Si \textsc{iv}  & 1394 & 8.89 \\
\hline
\end{tabular}
\end{table}

\subsection{Soft X-ray}

Current X-ray telescopes, i.e.\ \textit{Chandra}\footnote{http://chandra.harvard.edu/}, \textit{XMM-Newton}\footnote{http://xmm.esac.esa.int/}, and \textit{Suzaku}\footnote{http://www.jaxa.jp/projects/sat/astro\_e2/index\_e.html}, have already detected continuum and metal-line emission from the diffuse, intracluster gas. To map the halo gas out to larger radii and in lower mass haloes with future instruments, a combination of high angular and spectral resolution and a large field of view are preferred \citep[e.g.][]{Bertone2010a}. Higher spectral resolution results in higher signal-to-noise for unresolved lines. As the emission is dominated by bright, compact regions, high angular resolution is necessary to prevent smearing out the emission, in which case it would be harder to detect as well as impossible to correctly identify its origin. \textit{ASTRO-H}\footnote{http://astro-h.isas.jaxa.jp/} will be launched in a few years. While its spectral resolution is relatively good (5~eV), its field of view is rather small ($3\times3$ arcmin$^2$) and its spatial resolution is coarse (1.3~arcmin) \citep[e.g.][see also Table~\ref{tab:instr}]{Soong2011}. Therefore, it is not an ideal instrument for studying halo X-ray profiles, although it will do relatively well for large targets, such as massive clusters. A number of X-ray missions have been proposed in the past years, such as \textit{IXO/ATHENA}\footnote{
http://ixo.gsfc.nasa.gov/, http:/
/sci.esa.int/ixo} and \textit{ORIGIN} \citep{Herder2012}. Their proposed specifications are listed in Table~\ref{tab:instr}, as are those for existing and future UV and optical instruments, which are discussed in Sections~\ref{sec:UVlowz} and~\ref{sec:UVhighz}, respectively.

\begin{table*}
\caption{\label{tab:instr} \small Summary of the most relevant technical specifications for instruments discussed in the main text.}
\begin{tabular}[t]{llllll}
\hline
\hline \\[-3mm]
Telescope (instrument) & Field of view & Angular resolution & Spectral resolution & Effective area & Instrumental background \\
  & (arcmin$^2$) & (arcsec) &  & (cm$^2$) & (s$^{-1}$cm$^{-2}$keV$^{-1}$) \\
\hline \\[-4mm]
\textit{ORIGIN} (CRIS) & 10$\times$10 ($30\times30$) & 30 & 2.5 (5) eV & $10^3$ & 0.02 \\
\textit{IXO/ATHENA} (XMS) & $2\times$2 ($5\times5$) & 5 & 2.5 (10) eV & $3\times10^4$ & 0.02 \\
\textit{ASTRO-H} (SXS) & $3\times3$ & 78 & 5 eV & 50 & 0.002 \\
\textit{Suzaku} (XIS) & $18\times18$ & 120 & 70 eV & 100 & $2\times10$ \\ 
\hline
\textit{FIREBALL} & $2.67\times2.67$ & 10 & 0.4 \AA & $8\times10^4$ & \\
\hline
\textit{Hale} (CWI) & $1\times0.67$ & $2.5\times1$ & 0.12 \AA & $2\times10^5$ & \\
\textit{Keck} (KCWI) & $0.33\times(0.14-0.57)$ & $0.35-1.4$ & 0.03-0.6 \AA & $8\times10^5$ & \\
\textit{VLT} (MUSE) & $1\times1$ & 0.2 & 0.4 \AA & $5\times10^5$ & \\
\hline
\end{tabular}
\end{table*}

For CRIS on \textit{ORIGIN}, the instrumental background is $2\times10^{-2}$~photon s$^{-1}$cm$^{-2}$keV$^{-1}$, which corresponds to $1.3\times10^3$~photon s$^{-1}$sr$^{-1}$keV$^{-1}$ (with a pixel size of 300~$\mu$m and 24~arcsec). In the narrow-field mode, the energy resolution is 2.5~eV, so the instrumental background is $3$~photon s$^{-1}$sr$^{-1}$. For an effective area of $10^3$~cm$^2$, the surface brightness of the instrumental background thus equals $3\times10^{-3}$~photon s$^{-1}$cm$^{-2}$sr$^{-1}$. (It would be twice as high in the wide-field mode.)

For XMS on \textit{IXO/ATHENA} the instrumental background is also $2\times10^{-2}$~photon s$^{-1}$cm$^{-2}$keV$^{-1}$, which corresponds to $8.5\times10^4$~photon s$^{-1}$sr$^{-1}$keV$^{-1}$ (with a pixel size of 300~$\mu$m and 3~arcsec). For an energy resolution of 2.5 eV, which is appropriate for the narrow-field mode, this corresponds to $2\times10^2$~photon s$^{-1}$sr$^{-1}$. For an effective area of $3\times10^4$~cm$^2$, the surface brightness of the instrumental background thus equals $7\times10^{-3}$~photon s$^{-1}$cm$^{-2}$sr$^{-1}$. (It would be four times as high in the wide-field mode.)

For SXS on \textit{ASTRO-H} the instrumental background is an order of magnitude lower, $2\times10^{-3}$~photon s$^{-1}$cm$^{-2}$keV$^{-1}$, which corresponds to $7.0\times10^2$~photon s$^{-1}$sr$^{-1}$keV$^{-1}$ (with a pixel size of 830~$\mu$m and 29~arcsec). For an energy resolution of 5 eV, this corresponds to $3$~photon s$^{-1}$sr$^{-1}$. For an effective area of $50$~cm$^2$, the surface brightness of the instrumental background equals $7\times10^{-2}$~photon s$^{-1}$cm$^{-2}$sr$^{-1}$, an order of magnitude higher than for XMS on \textit{IXO/ATHENA}.

For XIS on \textit{Suzaku} the instrumental background is higher, $\sim2\times10$~photon s$^{-1}$cm$^{-2}$keV$^{-1}$ \citep{Tawa2008}, but because it contains many pixels, this corresponds to only $4.2\times10^2$~photon s$^{-1}$sr$^{-1}$keV$^{-1}$ (with a pixel size of 24~$\mu$m and 1~arcsec). Given an energy resolution of 70 eV, this amounts to 30~photon s$^{-1}$sr$^{-1}$. For an effective area of $100$~cm$^2$, the surface brightness of the instrumental background equals $3\times10^{-1}$~photon s$^{-1}$cm$^{-2}$sr$^{-1}$, a factor of four higher than for SXS on \textit{ASTRO-H}.

\citet{Hickox2006} determined that the unresolved cosmic X-ray background is about $14\pm1$~photon s$^{-1}$cm$^{-2}$sr$^{-1}$keV$^{-1}$ in the energy range we are interested in ($0.5-1$~keV). This unresolved signal is thought to be dominated by diffuse Galactic emission and local thermal-like emission. Part of this unresolved background could be resolved in deeper observations \citep{Hickox2007}. We will conservatively use the limit of 14~photon s$^{-1}$cm$^{-2}$sr$^{-1}$keV$^{-1}$ as it is an upper limit on the unresolved background for future observations. For the narrow-field mode, the energy resolution of CRIS on \textit{ORIGIN} and XMS on \textit{IXO/ATHENA} (i.e.\ 2.5~eV) this corresponds to about $3\times10^{-2}$~photon s$^{-1}$cm$^{-2}$sr$^{-1}$. The instrumental background is therefore lower than the unresolved X-ray background. Observations should be able to probe down to $10^{-1}$~photon s$^{-1}$cm$^{-2}$sr$^{-1}$ without being strongly affected by the background. For this surface brightness a 1~Msec exposure would result in about 2 detected photons per resolution element with \textit{ORIGIN} at 30~arcsec resolution and with \textit{IXO/ATHENA} at 5~arcsec resolution.

The energy resolutions of SXS on \textit{ASTRO-H} and of the wide-field mode for CRIS on \textit{ORIGIN} are two times worse than the narrow-field mode for CRIS and XMS. The unresolved cosmic X-ray background is therefore $7\times10^{-2}$~photon s$^{-1}$cm$^{-2}$sr$^{-1}$, comparable to the instrumental background of SXS on \textit{ASTRO-H}. The wide-field mode for XMS on \textit{IXO/ATHENA} leads to unresolved cosmic X-ray backgrounds of $14\times10^{-2}$~photon s$^{-1}$cm$^{-2}$sr$^{-1}$. Observations can therefore only probe down to several times $10^{-1}$~photon s$^{-1}$cm$^{-2}$sr$^{-1}$ without being strongly affected by the background. Assuming a detection limit of $3\times10^{-1}$~photon s$^{-1}$cm$^{-2}$sr$^{-1}$, a 1~Msec exposure would result in about 2 detected photons per resolution element on \textit{ASTRO-H} at 78~arcsec resolution.

The energy resolution of XIS on \textit{Suzaku} is only about 70 eV, as this is an imager and not a spectrograph. Therefore, the unresolved X-ray background corresponds to $1$~photon s$^{-1}$cm$^{-2}$sr$^{-1}$, a factor of three larger than its instrumental background. Observations can only probe down to about $5$~photon s$^{-1}$cm$^{-2}$sr$^{-1}$ without being strongly affected by the background. Additionally, the low energy resolution, 70 eV, may lead to strong projection effects and would make identifying spectral lines very difficult. 

For reference, 
\begin{multline}
10^{-19}~\mathrm{erg s^{-1} cm^{-2} arcsec^{-2}}\approx \\ 2.4\times\Big(\dfrac{\lambda}{10\AA}\Big)\Big(\dfrac{1+z}{1.125}\Big)~\mathrm{photon s^{-1} cm^{-2} sr^{-1}}.
\end{multline}

\subsection{Low-redshift UV}

To detect UV metal-line emission as a tracer of halo gas, instruments should ideally have a large field of view and a high spatial and angular resolution, as is also the case for X-ray emission \citep{Bertone2010b}. The emission will be dominated by relatively high-density material. With high spatial resolution, it will be possible to measure its clumpiness.

A surface brightness limit of order $10^{-18}$~erg s$^{-1}$cm$^{-2}$arcsec$^{-2}$, as is appropriate for the \textit{FIREBALL}\footnote{http://www.srl.caltech.edu/sal/fireball.html} balloon experiment \citep{Tuttle2008, Tuttle2010},  will only be sufficient to detect metal-line emission in the centres of massive haloes. The wavelength range of \textit{FIREBALL} (2000--2200~\AA) is such that it will only probe the UV metal lines at a somewhat higher redshift than shown in this paper ($z\approx0.35$ for C~\textsc{iv}) where the surface brightness is a bit lower. A detection limit of order $10^{-21}$~erg s$^{-1}$cm$^{-2}$arcsec$^{-2}$ is envisioned for the next generation UV mission \textit{ATLAST} \citep{Postman2009}, which would be a huge improvement.

For reference, 
\begin{multline}
10^{-19}~\mathrm{erg s^{-1} cm^{-2} arcsec^{-2}}\approx \\ 2.7\times10^2\Big(\dfrac{\lambda}{10^3\AA}\Big)\Big(\dfrac{1+z}{1.25}\Big)~\mathrm{photon s^{-1} cm^{-2} sr^{-1}}.
\end{multline}

\subsection{High-redshift UV}

It is possible to detect rest-frame UV emitted at high redshift from the ground, in the optical. In the near future, several integral field unit spectrographs will become operational, such as the Multi-Unit Spectroscopic Explorer\footnote{http://muse.univ-lyon1.fr/} \citep[MUSE;][]{Bacon2010}  on the \textit{Very Large Telescope} and the Keck Cosmic Web Imager\footnote{http://www.srl.caltech.edu/sal/keck-cosmic-web-imager.html} \citep[KCWI;][]{Martin2010} on \textit{Keck}. The Cosmic Web Imager\footnote{http://http://www.srl.caltech.edu/sal/cosmic-web-imager.html} \citep[CWI;][]{Rahman2006, Matuszewski2010} is already installed on the 200-inch \textit{Hale} telescope.

With narrow-band observations it has already been demonstrated that diffuse H\,\textsc{i} Ly$\alpha$ emission can be detected down to $\sim10^{-18}$~erg s$^{-1}$cm$^{-2}$arcsec$^{-2}$ in individual objects \citep[e.g.][]{Steidel2000, Matsuda2004} and down to $\sim10^{-19}$~erg s$^{-1}$cm$^{-2}$arcsec$^{-2}$ in stacks \citep{Steidel2011}. CWI is expected to reach similar depths. New instrumentation, such as MUSE and KCWI, will allow us to go one order of magnitude deeper. This increase in sensitivity is essential for the detection of metal lines \citep{Bertone2012}.

\section{Surface brightness profiles} \label{sec:SB}

\begin{figure*}
\center
\includegraphics[scale=.8]{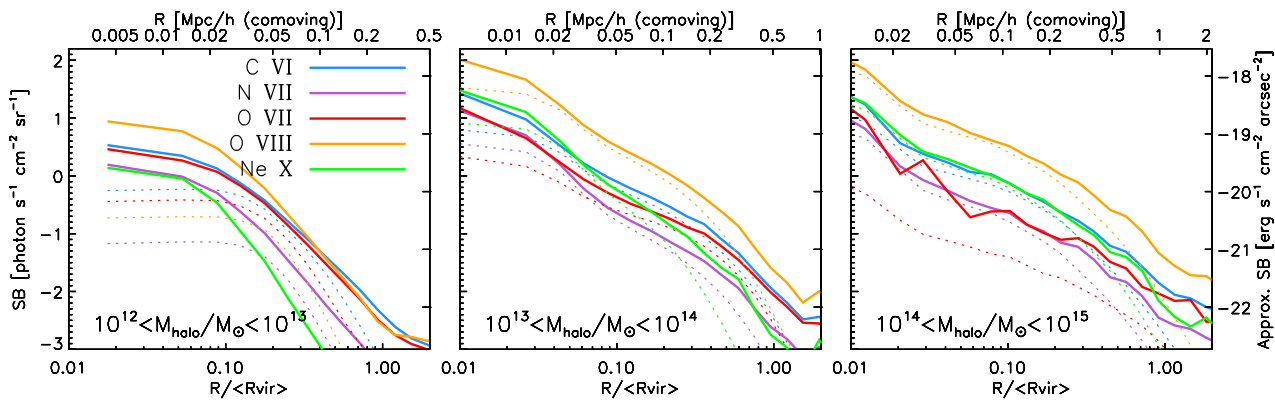}
\caption {\label{fig:profileXrayz0p125} Mean (solid curves) and median (dotted curves) surface brightness as a function of radius at $z=0.125$ for the soft X-ray lines indicated in the legend. We use three different halo mass bins with masses of $M_\mathrm{halo}=10^{12-13}$~M$_\odot$ (left), $10^{13-14}$~M$_\odot$ (middle), and $10^{14-15}$~M$_\odot$ (right). The thickness of the slices is 1.7, 3.7, and 5.6~comoving~$h^{-1}$Mpc, respectively, corresponding to four times the largest virial radius in each mass bin. The radius on the bottom $x$-axis is normalized by dividing by the median virial radius in each mass bin. The surface brightness in photon~s$^{-1}$cm$^{-2}$sr$^{-1}$ indicated by the left $y$-axis is exact. For the right $y$-axis it has been converted to erg~s$^{-1}$cm$^{-2}$arcsec$^{-2}$ using $\langle\lambda\rangle=22.2$~\AA~(0.558~keV). The pixel size was 5~arcsec (8.9~comoving~$h^{-1}$kpc; 7.9~proper~$h^{-1}$kpc) before binning. O\,\textsc{viii} is the brightest line at all halo masses, followed by C\,\textsc{vi}. The relative strengths of the lines vary with halo mass. The profiles flatten at $R\lesssim10$~$h^{-1}$kpc, because this region is dominated by the ISM, which we excluded from our analysis.}
\end{figure*}

For all the emission lines listed in Table~\ref{tab:lines}, we calculate surface brightness maps for each halo and stack the maps for haloes within a certain halo mass range. We then average the mean surface brightness maps azimuthally to make emission profiles. In this Section we discuss the surface brightness profiles and the detectability of the different emission lines. Median profiles are calculated by taking the median of the pixels in the stacked images and then averaging those azimuthally.

\subsection{Soft X-ray}

In Figure~\ref{fig:profileXrayz0p125} we show the mean (median) X-ray line surface brightness profiles for haloes at $z=0.125$ as solid (dotted) curves. We use three halo mass bins, roughly corresponding to (the haloes of) galaxies, groups, and clusters, with masses of $M_\mathrm{halo}=10^{12-13}$~M$_\odot$, $10^{13-14}$~M$_\odot$, and $10^{14-15}$~M$_\odot$. We show the profiles out to twice the median virial radius and with a thickness of four times the maximum virial radius in each mass bin. A spectral resolution of 2.5~eV corresponds to a thickness of about 15~Mpc. Using 15~Mpc thick slices would only mildly flatten the profiles of the lowest mass haloes around $R_\mathrm{vir}$, which is below the detection limit of proposed instruments.

The strongest observable soft X-ray line is O~\textsc{viii} (0.654 keV). It is followed by C~\textsc{vi} (0.367 keV) and O~\textsc{vii} (0.561 keV) for galaxies. Ne\,\textsc{x} (1.021 keV) has the lowest surface brightness for $M_\mathrm{halo}=10^{12-13}$~M$_\odot$. In centres of groups and clusters, Ne~\textsc{x} is the second strongest line, closely followed by C~\textsc{vi}. 
As mentioned in Section~\ref{sec:em}, O~\textsc{vii} is a triplet and the emission will be about a factor of two stronger when the contributions of the three lines are added.
For both galaxies and groups, the Ne~\textsc{x} profile is steeper than that of all the other lines shown. N~\textsc{vii} (0.500 keV) is similar to O~\textsc{vii} for cluster haloes. We also computed the emissivities of C~\textsc{v} (0.308 keV), N~\textsc{vi} (0.420 keV), Ne~\textsc{ix} (0.922 keV), Mg~\textsc{xii} (1.472 keV), Si~\textsc{xiii} (1.865 keV), Si~\textsc{xv} (2.460 keV), and Fe~\textsc{xvii} (0.727 keV), which were in general weaker than the lines shown \citep[see also][]{Bertone2010a}. The surface brightnesses of Ne~\textsc{ix} and Fe~\textsc{xvii} are as strong as those of N~\textsc{x} in galaxies and Fe~\textsc{xvii} is as high as O~\textsc{vii} for $R<0.2R_\mathrm{vir}$ in groups. Both lines are fainter in cluster centres, but possibly still detectable.

The median surface brightness profiles are lower than the mean profiles at all radii. A large difference between mean and median profiles indicates that the emission is dominated by relatively rare, bright pixels and thus that the emission is clumpy. For $M_\mathrm{halo}=10^{12-13}$~M$_\odot$, Ne~\textsc{x} and O~\textsc{viii} exhibit large differences between mean and median profile (the median Ne~\textsc{x} profile is lower than the range shown), whereas for $M_\mathrm{halo}=10^{14-15}$~M$_\odot$ this is only the case for O~\textsc{vii}. The difference increases at very small radii ($R<0.1R_\mathrm{vir}$) for $M_\mathrm{halo}=10^{12-13}$~M$_\odot$ and towards the edge of the halo ($R>0.5R_\mathrm{vir}$) for all haloes. At intermediate radii the difference between the mean and median profiles is small for the majority of the lines, implying that the X-ray emission is relatively smooth.

At a limiting surface brightness of $10^{-1}$~photon s$^{-1}$cm$^{-2}$sr$^{-1}$, as is appropriate for CRIS on \textit{ORIGIN} and XMS on \textit{IXO/ATHENA}, we would be able to detect O~\textsc{viii} emission out to 80 per cent of the virial radius of a cluster with $M_\mathrm{halo}=10^{14-15}$~M$_\odot$, C~\textsc{vi} out to 40 per cent, and O~\textsc{vii} and N~\textsc{vii} out to 20 per cent. For groups, $M_\mathrm{halo}=10^{13-14}$~M$_\odot$, O~\textsc{vii} can be observed out to the same physical scale, $\sim$200~comoving~$h^{-1}$kpc, which corresponds to a larger fraction of the virial radius, 0.4$R_\mathrm{vir}$. O~\textsc{viii} is observable out to 0.7$R_\mathrm{vir}$ and C~\textsc{vi} out to 0.5$R_\mathrm{vir}$. For $M_\mathrm{halo}=10^{12-13}$~M$_\odot$, O~\textsc{viii}, O~\textsc{vii}, and C~\textsc{vi} can be observed out to the same radius of 0.3$R_\mathrm{vir}$.

SXS on \textit{ASTRO-H} has poor angular resolution (78~arcsec which equals 0.14~comoving~$h^{-1}$Mpc), but it may be able to detect O\,\textsc{viii} emission in the centres of groups and clusters. It will not be able to see detailed structure, because the emission would be contained in a single pixel.

\subsection{Low-redshift UV}

\begin{figure*}
\center
\includegraphics[scale=.8]{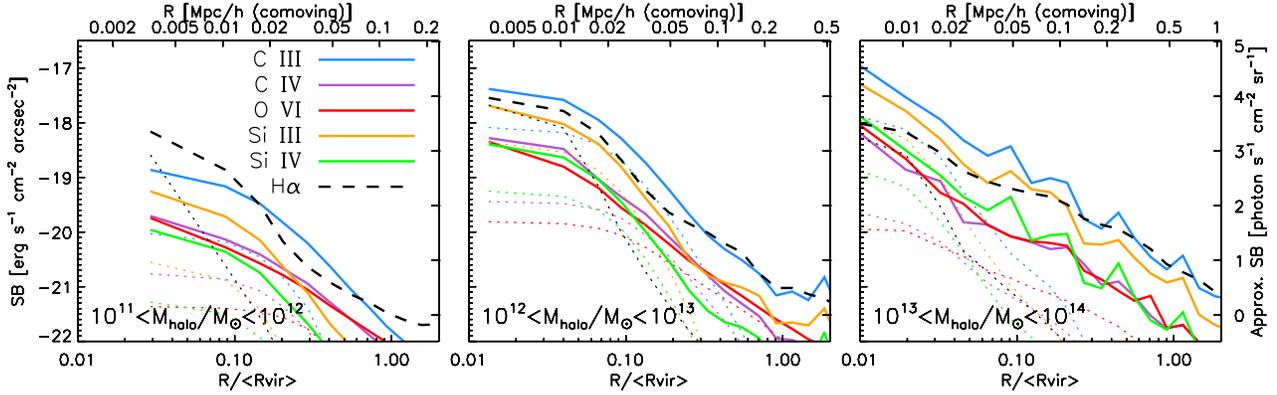}
\caption {\label{fig:profileUVz0p25} Mean (solid curves) and median
(dotted curves) surface brightness profiles for the emission lines and halo masses indicated in the legend at $z=0.25$. Note that the halo masses, $M_\mathrm{halo}=10^{11-12}$~M$_\odot$ (left), $10^{12-13}$~M$_\odot$ (middle), and $10^{13-14}$~M$_\odot$ (right), are lower than in Figure~\ref{fig:profileXrayz0p125}. The dashed curve shows the median surface brightness profile for H$\alpha$. The thickness of the slices is 0.8, 1.7, and 3.3~comoving~$h^{-1}$Mpc, respectively, corresponding to four times the largest virial radius in each mass bin. The surface brightness in erg~s$^{-1}$cm$^{-2}$arcsec$^{-2}$ on the left $y$-axis is exact. For the right $y$-axis, it has been converted to photon~s$^{-1}$cm$^{-2}$sr$^{-1}$ using $\langle\lambda\rangle=1232$~\AA. The pixel size was 2~arcsec (3.5~comoving~$h^{-1}$kpc; 2.8~proper~$h^{-1}$kpc) before binning. C\,\textsc{iii} is the brightest UV metal line, followed by Si\,\textsc{iii}. }
\end{figure*}

Figure~\ref{fig:profileUVz0p25} shows the surface brightness for the UV metal lines listed in Table~\ref{tab:lines} at $z=0.25$ for haloes with masses of $M_\mathrm{halo}=10^{11-12}$~M$_\odot$, $10^{12-13}$~M$_\odot$, and $10^{13-14}$~M$_\odot$. Line styles are identical to those in Figure~\ref{fig:profileXrayz0p125} (i.e.\ solid indicates means, dotted medians), but colours show different emission lines, as indicated in the legend. We also include hydrogen H$\alpha$ (black, dashed curves) from the same haloes and discuss it in Section~\ref{sec:Ha}. 

C~\textsc{iii} (977 \AA) is the brightest line, followed by Si~\textsc{iii} (1207 \AA). The emission profile for all lines peaks in the halo core, where it may, however, be outshone by the central galaxy (recall that we omitted emission from gas with n$_\mathrm{H}>0.1$~cm$^{-3}$) O~\textsc{vi} (1032 \AA) gives the most extended profile, in the sense that the ratio between emission in the halo core and outskirts is smallest. This is caused by the fact that O~\textsc{vi} traces hotter and more diffuse gas than the other lines. C~\textsc{iv} (1548 \AA) and Si~\textsc{iv} (1294 \AA) are weaker than C~\textsc{iii} and Si~\textsc{iii}, but because they have frequencies redward of Ly$\alpha$, they are less easily absorbed by intervening gas clouds. We also computed the emissivities of He~\textsc{ii} (1640~\AA)~and N~\textsc{v} (1239~\AA), which were weaker and therefore not shown \citep[see also][]{Bertone2010b}.

The median surface brightness profiles are lower than the mean profiles. For $M_\mathrm{halo}=10^{11-12}$~M$_\odot$, all lines show a large difference between mean and median profile, especially in the inner parts. For $M_\mathrm{halo}=10^{12-13}$~M$_\odot$ the difference is very small at $0.05R_\mathrm{vir}<R<0.4R_\mathrm{vir}$. For $M_\mathrm{halo}=10^{13-14}$~M$_\odot$, the O~\textsc{vi} clumpiness decreases with radius, whereas the clumpiness of the other lines increases. For $M_\mathrm{halo}>10^{12}$~M$_\odot$, O~\textsc{vi} exhibits the smallest difference between mean and median profiles, consistent with the expectation that it traces hotter, more diffuse gas.

A surface brightness limit of $10^{-18}$~erg s$^{-1}$cm$^{-2}$arcsec$^{-2}$ should enable us to observe metal-line emission from the centres of fairly massive galaxies ($M_\mathrm{halo}>10^{12}$~M$_\odot$). However, it is possible that the central galaxy will outshine the halo in the halo cores. 

By stacking haloes, \textit{FIREBALL} could in principle probe down to a surface brightness limit of $10^{-19}$~erg s$^{-1}$cm$^{-2}$arcsec$^{-2}$. Its wavelength rage is such that it can only detect C\,\textsc{iii} and Si\,\textsc{iii} at $z\approx1$. As the surface brightness is $\sim0.3$~dex lower at $z=1$ as compared to $z=0.25$ (not shown), it may see emission out to $\sim0.1R_\mathrm{vir}$ for $M_\mathrm{halo}>10^{12}$~M$_\odot$. A complicating factor is that C\,\textsc{iii}, O\,\textsc{vi} and Si\,\textsc{iii} have frequencies blueward of Ly$\alpha$ and their light could therefore be absorbed by intervening hydrogen, an effect that is largest for C\,\textsc{iii} and smallest for Si\,\textsc{iii}.

With a detection limit of order $10^{-21}$~erg s$^{-1}$cm$^{-2}$arcsec$^{-2}$ we would be able to detect C\,\textsc{iii} out to the virial radius and several other metal lines out to 50 per cent of $R_\mathrm{vir}$. Such a low surface brightness limit may be attainable for the next generation UV mission, \textit{ATLAST} \citep{Postman2009}. As Ly$\alpha$ emission is even brighter, such a mission can even probe the gas in emission outside galactic haloes, in the filaments of the cosmic web \citep[e.g.][]{Furlanetto2003}.

\subsection{High-redshift UV} \label{sec:UV}

\begin{figure*}
\center
\includegraphics[scale=.8]{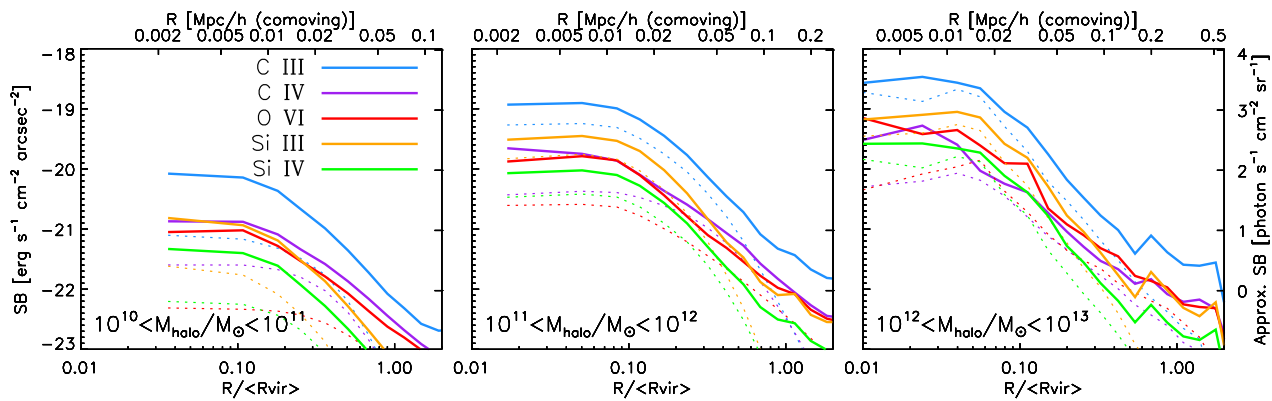}
\caption {\label{fig:profileUVz3p0} Mean (solid curves) and median
(dotted curves) surface brightness profiles for the rest-frame UV emission lines and halo masses indicated in the legend at $z=3$. Note that the halo masses, $M_\mathrm{halo}=10^{10-11}$~M$_\odot$ (left), $10^{11-12}$~M$_\odot$ (middle), and $10^{12-13}$~M$_\odot$ (right), are lower than in Figures~\ref{fig:profileXrayz0p125} and~\ref{fig:profileUVz0p25}. The thickness of the slices is 0.45, 0.95, and 1.25~comoving~$h^{-1}$Mpc, respectively, corresponding to four times the largest virial radius in each mass bin. The pixel size was 0.2~arcsec (4.6~comoving~$h^{-1}$kpc; 1.2~proper~$h^{-1}$kpc) before binning. C\,\textsc{iii} is the brightest line. C\,\textsc{iv}, Si\,\textsc{iii}, and Si\,\textsc{iv} are of similar strength. The surface brightnesses are lower than at $z=0.25$ and the flattening at small radii is more pronounced.}
\end{figure*}

Figure~\ref{fig:profileUVz3p0} shows the UV line surface brightness profiles at $z=3$ for haloes with masses of $M_\mathrm{halo}=10^{10-11}$~M$_\odot$, $10^{11-12}$~M$_\odot$, and $10^{12-13}$~M$_\odot$. The surface brightness is about an order of magnitude lower at $z=3$ than at $z=0.25$ at a fixed fraction of $R_\mathrm{vir}$. Additionally, the profile shows a strong flattening inside 0.1$R_\mathrm{vir}$. This is because we are probing down to smaller physical scales, where the emission is dominated by star-forming gas, which we have excluded from our calculations.

Comparing our metal-line results to the Ly$\alpha$ profile observed by \citet{Steidel2011} for $M_\mathrm{halo}\sim10^{12}$~M$_\odot$ and $\langle z\rangle=2.65$, we predict that the C~\textsc{iii} surface brightness is about 1--2~dex lower than the Ly$\alpha$ brightness (ignoring absorption by intervening gas). The other metal lines that are shown are about 0.5--1~dex fainter than C~\textsc{iii}. 

Because C~\textsc{iii} has a much shorter wavelength than Ly$\alpha$, it is strongly absorbed by the intervening medium. It is therefore possible that the other metal lines will be observed to be stronger than  C~\textsc{iii}. We used the observed effective  Ly$\alpha$ optical depth, $\tau_\mathrm{eff}$, from \citet{Schaye2003} to estimate the average attenuation of C~\textsc{iii} emission at $z=3$. This line is redshifted to the  H~\textsc{i} Ly$\alpha$ wavelength ($\lambda_\mathrm{Ly\alpha}=1216$~\AA) at z=2.21 and to the H~\textsc{i} Ly$\beta$ wavelength ($\lambda_\mathrm{Ly\beta}=1026$ \AA) at z=2.38. At these redshifts, $\tau_\mathrm{eff}=0.17$ and 0.20, respectively. Using these two hydrogen lines, which dominate the absorption, we obtained the average C~\textsc{iii} optical depth $\tau_\mathrm{C\textsc{iii}} = \tau_\mathrm{eff}^{z=2.21} + \dfrac{\lambda_\mathrm{Ly\beta}f_\mathrm{Ly\beta}}{\lambda_\mathrm{Ly\alpha}f_\mathrm{Ly\alpha}}\tau_\mathrm{eff}^{z=2.38}=0.20$, where $f_\mathrm{Ly\alpha}=0.416$ and $f_\mathrm{Ly\beta}=0.079$ are the H~\textsc{i} Ly$\alpha$ and Ly$\beta$ oscillator strenghts. On average, only a fraction of $1-e^{-\tau_\mathrm{C\textsc{iii}}}=0.18$ of the C~\textsc{iii} emission at $z=3$ is absorbed. We therefore predict that intergalactic absorption by hydrogen will not change the $z=3$ surface brightness profiles significantly. 

We calculated, but do not show, the Ly$\alpha$ surface brightness profile from our simulations in the optically thin limit and without a contribution from star formation. The observed profile is $\sim0.7$~dex higher than the predicted profile, but this is not surprising considering that most of the Ly$\alpha$ emission is thought to be originating from H~\textsc{ii} regions in the ISM \citep[e.g.][]{Furlanetto2005, Steidel2011} and that we excluded this gas from our analysis. 

For $M_\mathrm{halo}=10^{10-11}$~M$_\odot$, all lines show a 0.3--1~dex difference between mean and median profiles. The difference is largest in the inner parts. For $M_\mathrm{halo}>10^{11}$~M$_\odot$ at intermediate radii ($0.03R_\mathrm{vir}<R<0.6R_\mathrm{vir}$) the UV emission is least clumpy.

A detection limit of $10^{-19}$~erg s$^{-1}$cm$^{-2}$arcsec$^{-2}$, which should be possible for CWI, is not low enough for the detection of most metal lines. C\,\textsc{iii} emission could be seen out to almost 0.1~$R_\mathrm{vir}$ for massive haloes, if it is not attenuated by the IGM. With a detection limit of $10^{-20}$~erg s$^{-1}$cm$^{-2}$arcsec$^{-2}$, which should be possible for MUSE and KCWI when stacking many galaxies, C\,\textsc{iii} can be seen out to 0.3$R_\mathrm{vir}$ for $M_\mathrm{halo}=10^{11-12}$~M$_\odot$ and out to 0.2$R_\mathrm{vir}$ for $10^{12-13}$~M$_\odot$ at $z=3$. Note, however, that MUSE can only see C\,\textsc{iii} emission for $z>3.8$. Other metal lines are weaker, but can still be probed to 10 per cent of $R_\mathrm{vir}$ for $M_\mathrm{halo}>10^{11}$~M$_\odot$. The emission may, however, in reality be brighter at these small radii, because we excluded emission from gas with $n_\mathrm{H}>0.1$~cm$^{-3}$. We predict that the surface brightness for lower-mass haloes is below this detection limit.

\subsection{Low-redshift H$\alpha$} \label{sec:Ha}

H$\alpha$ (6563 \AA) emission at low redshift is observable with the same instruments as high redshift rest-frame UV emission. The predicted emission from gas around galaxies is shown in Figure~\ref{fig:profileUVz0p25} as a dashed, black curve. Interestingly, the H$\alpha$ emission is comparable to C~\textsc{iii} emission at $R>50$~kpc for all haloes with $M_\mathrm{halo}=10^{11-14}$~M$_\odot$. However, for $R<50$~kpc, H$\alpha$ is much brighter than C~\textsc{iii} for the lowest halo mass bin ($M_\mathrm{halo}=10^{11-12}$~M$_\odot$) and up to an order of magnitude fainter for the highest halo mass bin ($M_\mathrm{halo}=10^{13-14}$~M$_\odot$). We stress again that we ignored emission from gas with $n_\mathrm{H}>0.1$~cm$^{-3}$, implying that the true emission may be brighter. 

For $M_\mathrm{halo}=10^{11-12}$~M$_\odot$ the median surface brightness profile is significantly lower than the mean profile (2~dex at $0.2R_\mathrm{vir}$), but this difference decreases towards smaller radii. This indicates that the emission is more clumpy in the outer than in the inner halo. For higher-mass haloes, the difference is smaller, but also increases at large radii.

Assuming a limiting surface brightness of $10^{-20}$~erg s$^{-1}$cm$^{-2}$arcsec$^{-2}$, H$\alpha$ emission from halo gas can be detected out to $0.2R_\mathrm{vir}$ for haloes with $M_\mathrm{halo}=10^{11-12}$~M$_\odot$, to $0.3R_\mathrm{vir}$ for haloes with $M_\mathrm{halo}=10^{12-13}$~M$_\odot$, and out to $0.6R_\mathrm{vir}$ for group-sized haloes. This may be feasible when stacking deep observations centred on galaxies with MUSE or KCWI.

\section{Flux-weighted physical properties} \label{sec:prop}

\begin{figure*}
\center
\includegraphics[scale=.6]{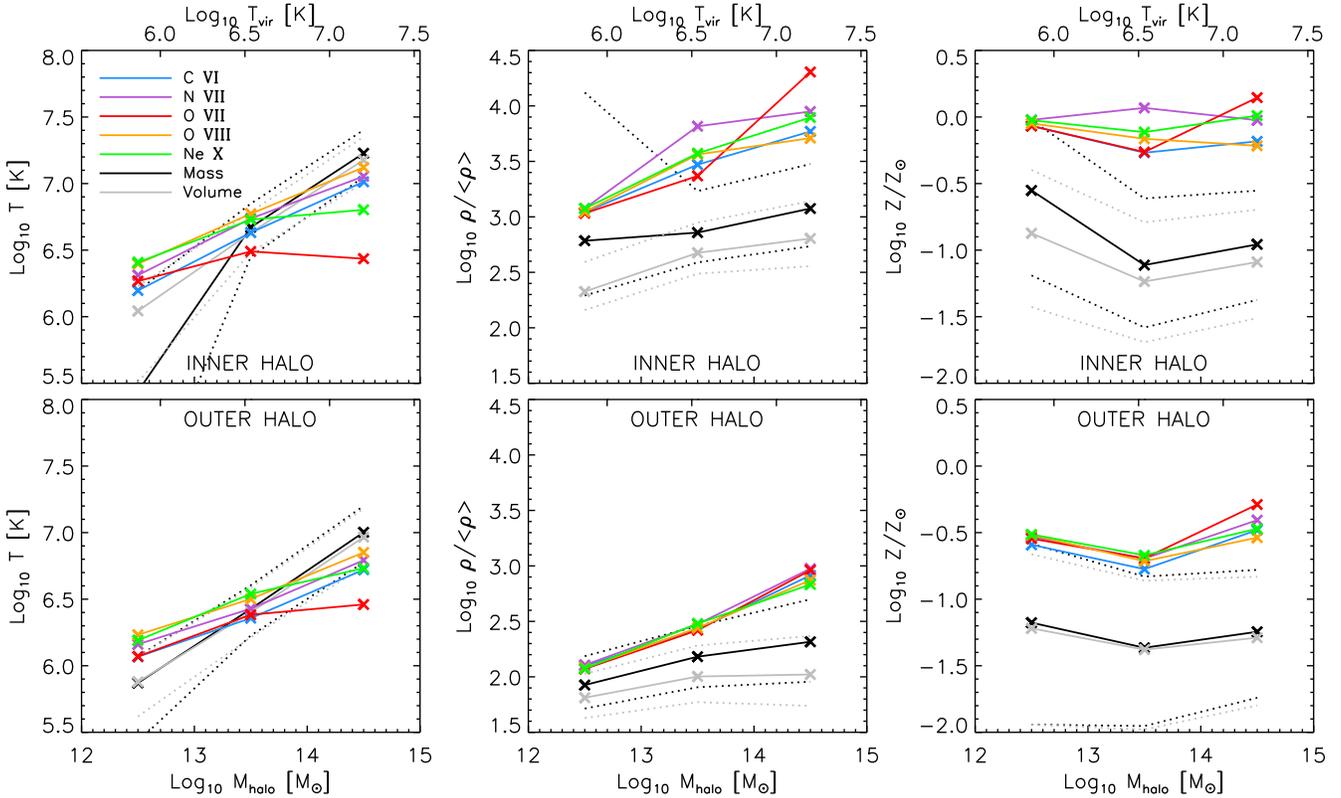}
\caption {\label{fig:propXrayz0p125} Mass-weighted (black curves), volume weighted (grey curves) and flux-weighted (coloured curves for different soft X-ray emission lines as indicated in the legend) median temperature, overdensity, and metallicity at $z=0.125$ as a function of halo mass. The top panels show the results for gas at $R<0.5R_\mathrm{vir}$, while the bottom panels show $0.5R_\mathrm{vir}-R_\mathrm{vir}$. The dotted curves show the mass- and volume-weighted 16th and 84th percentiles. The flux-weighted properties are biased towards high density and metallicity. The flux-weighted temperature is biased towards the temperature at which the line reaches its peak emissivity.}
\end{figure*}

The bright emission is dominated by collisionally ionized gas \citep{Bertone2010a, Bertone2010b, Bertone2012}. Most metal-line photons are emitted following excitation resulting from collisions between metal ions and free electrons. Hence, the emissivity is proportional to the square of the density and to the metallicity. In addition, the emissivity is strongly peaked at the temperature corresponding to the energy of the excited state. Hence, we expect the metal-line emission to be biased towards higher densities, towards higher metallicities, and towards temperatures near the peak of the emissivity curve. If the difference between the virial temperature and the temperature for which the emissivity peaks is large for a specific emission line and halo mass, then we may expect the flux-weighted density, temperature, and metallicity to be substantially biased relative to the typical physical conditions in the halo.

\begin{figure*}
\center
\includegraphics[scale=.6]{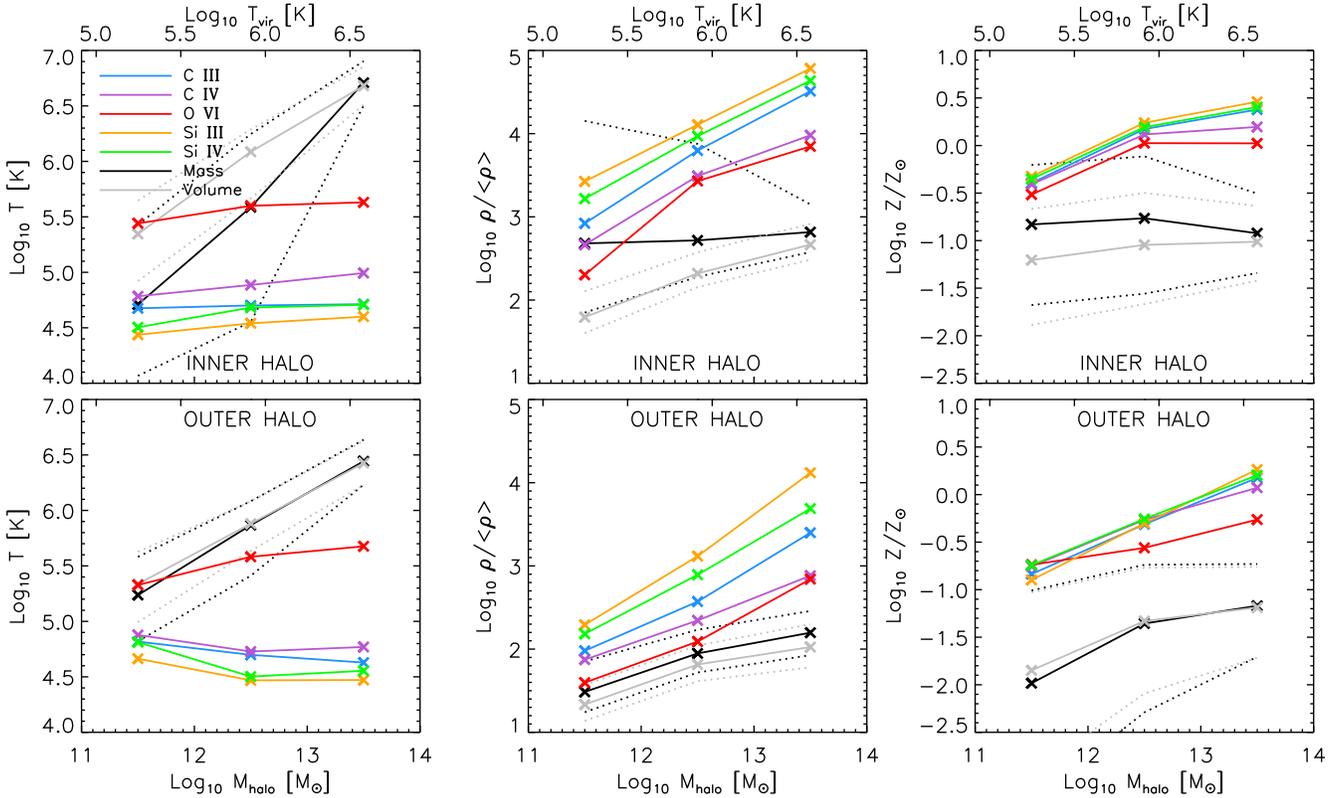}
\caption {\label{fig:propUVz0p25} Same properties as shown in Figure~\ref{fig:propXrayz0p125}, but for UV emission lines at $z=0.25$. Note that the axis ranges differ from Figures~\ref{fig:propXrayz0p125} and~\ref{fig:propUVz0p25}.}
\end{figure*}

To investigate possible observational biases when deriving physical properties of the halo gas, Figures~\ref{fig:propXrayz0p125}--\ref{fig:propUVz3p0} show the logarithms of the flux-weighted median temperature, overdensity, and metallicity as coloured curves for the same redshifts and halo mass bins as in Figures~\ref{fig:profileXrayz0p125}--\ref{fig:profileUVz3p0}. The legends indicate the different emission lines. Mass- and volume-weighted properties are shown as black and grey curves, respectively. The dotted curves show the 16th and 84th percentiles and the solid curves show medians. The top panels show the results at $R<0.5R_\mathrm{vir}$, while the bottom panels show $0.5R_\mathrm{vir}-R_\mathrm{vir}$. These radii are the true three-dimensional radii, but the results change by less than 0.1~dex if we use projected radii (not shown). 

\subsection{Soft X-ray}

Figure~\ref{fig:propXrayz0p125} shows flux-weighted properties at $z=0.125$ for soft X-ray lines.
The temperature at which the emissivity peaks is highest for Ne~\textsc{x} ($\approx10^{6.8}$~K) and O\,\textsc{viii} ($\approx10^{6.5}$~K), and it decreases slowly towards higher temperatures \citep[e.g. Figure~1 of][]{Bertone2010a}. This is the reason why, for galaxies and groups, these lines have the highest flux-weighted temperatures. For clusters, the Ne~\textsc{x}-weighted temperature is lower than that of O\,\textsc{viii} and close to the temperature where its emissivity peaks. The flux-weighted median temperatures increase less steeply with halo mass than the mass-weighted median temperature, although they still increase by $\sim0.2-0.9$~dex for an increase of 2~dex in halo mass. For the lowest (highest) mass bin the flux-weighted temperatures are higher (lower) than the mass-weighted temperature. The O\,\textsc{vii} flux-weighted temperature stays roughly constant with halo mass. This is due to the fact that the emissivity curve of O\,\textsc{vii} is more strongly peaked in temperature than the emissivity curves of the other lines \citep[see Figure~1 of][]{Bertone2010a}. It 
peaks 
at $10^{6.3}$~K and drops quickly for both higher and lower temperatures.

In the halo outskirts, the temperatures probed are somewhat lower, by $0.2-0.3$~dex, than in the inner parts. This is true for flux-weighted as well as mass-weighted median temperatures and densities, so they follow the same trend.

The densities are $0.5-0.7$~dex lower in the outer halo than in the inner halo. The flux-weighted overdensities of the gas responsible for the different lines are very similar and close to the 84th percentile of the mass-weighted mean overdensity which is $\sim0.2-0.7$~dex above the median mass-weighted density. The reason for this is that the emissivity scales as $\rho^2$, which biases the emission towards high densities. The flux-weighted densities increase by $\sim0.8$~dex over two orders of magnitude in halo mass, but the volume- and mass-weighted median densities increase by only 0.2--0.5~dex over the same range of halo masses.

The median mass-weighted metallicity is 0.04--0.2$Z_\odot$, but the flux-weighted metallicities are about 0.5-1.0~dex higher. For galaxies and groups, the median flux-weighted metallicities are similar to the 84th percentile of the mass-weighted metallicity. The flux-weighted metallicities for clusters are significantly above the 84th percentile, especially for O~\textsc{vii}, which also has the highest overdensity and lowest temperature. Thus, not surprisingly, X-ray metal-line emission is biased towards high-metallicity gas. As metal lines dominate the emission in the soft X-ray band (0.5-1.0~keV), the same will be true for broad band emission \citep[e.g.][]{Crain2010a}.

For clusters, we also find significant bias for line-of-sight velocities and velocity dispersions as measured with the soft X-ray lines described above (not shown). The velocity bias is highest for O~\textsc{vii}, which is also most discrepant in other properties. For harder X-ray lines, such as Fe~\textsc{xxv}, whose emissivity curves peak at higher temperatures, the bias for all quantities decreases and almost disappears for the flux-weighted velocities and velocity dispersions. We therefore conclude that it is essential to use metal lines with peak emissivity temperatures close to the virial temperature of the halo when deriving gas properties. We caution, however, that he bias does not vanish completely, so it is important to compare properties derived from real and simulated observations.

\subsection{Low-redshift UV} \label{sec:UVlowz}

Figure~\ref{fig:propUVz0p25} shows flux-weighted properties at $z=0.25$ for UV lines.
The flux-weighted temperature depends on the specific metal line. They are insensitive to the mass-weighted temperature of the halo, because the emissivities are strongly peaked at specific temperatures \citep{Bertone2010b}. Hence, detection of rest-frame UV metal-line emission puts strong constraints on the temperature of the emitting gas. The difference between Si\,\textsc{iii} and O\,\textsc{vi} is just over an order of magnitude. All UV metal-lines probe gas with temperatures 10$^{4.5-5}$~K with the exception of O\,\textsc{vi}, which is dominated by 10$^{5.5}$~K gas.

\begin{figure*}
\center
\includegraphics[scale=.6]{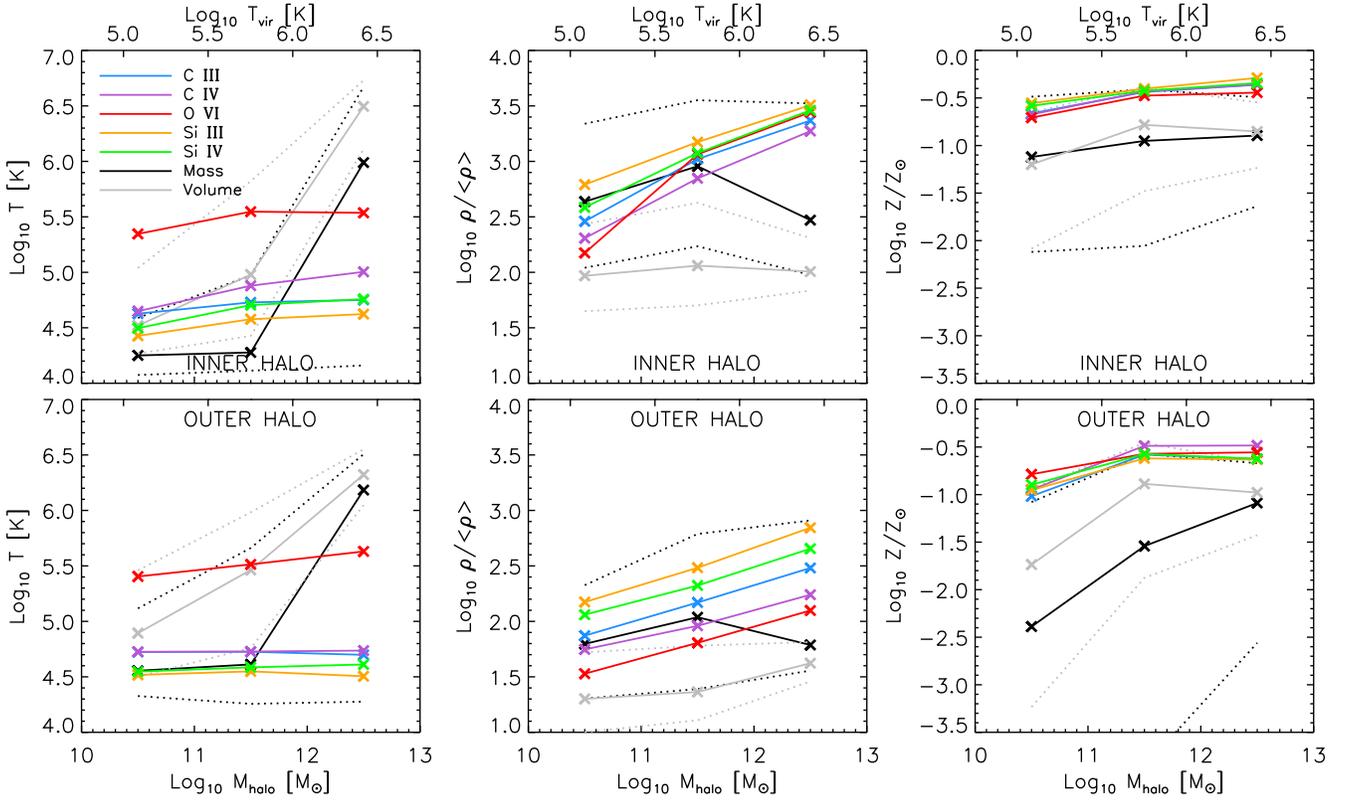}
\caption {\label{fig:propUVz3p0} Same properties as shown in Figure~\ref{fig:propXrayz0p125}, but for rest-frame UV emission lines at $z=3$. Note that the axis ranges differ from Figure~\ref{fig:propXrayz0p125}.}
\end{figure*}

The different metal lines probe very different densities, all higher than the volume-weighted overdensity. Because colder gas has higher density, the flux-weighted density increases as the corresponding temperature decreases. O\,\textsc{vi} results in the lowest flux-weighted overdensity. The flux-weighted density increases via C\,\textsc{iv} to C\,\textsc{iii} to Si\,\textsc{iv} and is highest for Si\,\textsc{iii}. For all halo masses, the silicon lines have higher flux-weighted densities than the 84th percentile of the mass-weighted distribution. In all haloes the flux-weighted median overdensities are higher than the volume-weighted 84th percentile. In group-sized haloes all lines are dominated by gas even denser than the mass-weighted 84th percentile. The difference between the O\,\textsc{vi} and Si\,\textsc{iii} flux-weighted overdensities is $\sim1$~dex. The flux-weighted overdensities increase more steeply with halo mass than the volume- and mass-weighted overdensity, by about 1.5~dex for an increase of 2~dex in halo mass. The overdensity is consistently lower by $0.3-0.8$~dex in the halo outskirts.

The flux-weighted median metallicity is 0.5--1.5~dex higher than the mass-weighted median metallicity and it is also higher than the mass-weighted 84th percentile. The discrepancy is larger in the halo outskirts. For high-mass haloes with $M_\mathrm{halo}>10^{12}$~M$_\odot$, O~\textsc{vi} probes gas with $\sim0.3$~dex lower metallicities than the other lines. In the inner parts of $M_\mathrm{halo}>10^{12}$~M$_\odot$ haloes, the emission of all lines is dominated by gas with supersolar metallicities, even though on average the gas is only enriched to $Z\approx0.1$~Z$_\odot$. All metallicities increase with halo mass. The metallicity difference between inner and outer halo is largest for the lowest-mass haloes.

\subsection{High-redshift UV} \label{sec:UVhighz}

In Figure~\ref{fig:propUVz3p0} the flux-weighted properties are shown at $z=3$ for UV lines. The general picture is unchanged at high redshift as compared to low redshift (see Figure~\ref{fig:propUVz0p25}).
The flux-weighted temperatures are insensitive to halo mass. They trace temperatures near or above the median mass-weighted temperature for low-mass haloes ($M_\mathrm{halo}<10^{12}$~M$_\odot$ ), but the opposite is true for higher-mass haloes. The scatter in temperature increases with halo mass and becomes very large, with the mass-weighted 84th percentile 2.5~dex above the 16th percentile. This reflects the bimodal nature of halo gas in massive high-redshift galaxies, where the gas either has a temperature close to the virial temperature or has $T<10^5$~K \citep[e.g.][]{VoortSchaye2012}. O~\textsc{vi} traces gas about an order of magnitude warmer than the other metal lines shown, because its emissivity peaks around $10^{5.5}$~K compared with $10^{4.5-5}$~K for the other metal lines \citep[see Figure~1 of][]{Bertone2012}.

\begin{figure*}
\center
\includegraphics[scale=.8]{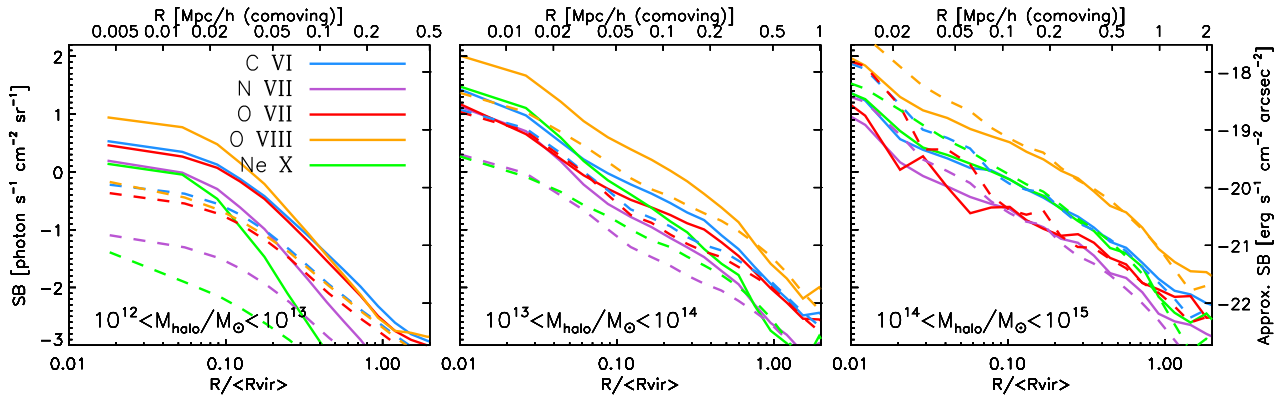}
\caption {\label{fig:AGNprofileXrayz0p125} Surface brightness profiles as in Figure~\ref{fig:profileXrayz0p125} for a simulation without AGN feedback (solid curves, identical to those in Figure~\ref{fig:profileXrayz0p125}) and for a simulation with AGN feedback (dashed curves).}
\end{figure*}

The density differences between different lines is somewhat reduced at high redshift as compared to low redshift. The flux-weighted overdensities are lower, but the corresponding physical densities are higher at high redshift.

At $z=3$ the scatter in the density is larger than at $z=0.25$: the difference between the 16th and 84th percentile is more than an order of magnitude as opposed to $0.4$~dex at $z=0.25$. As a result, none of the lines are dominated by overdensities above the 84th percentile at high redshift. The flux-weighted densities increase with halo mass, even if the volume- and mass-weighted densities do not. The median densities are lower in the outskirts than in the centre, by up to $\sim0.6$~dex.

As at low redshifts, the flux-weighted properties are biased towards high metallicities. All lines trace gas with metallicities of about 0.1--0.5~Z$_\odot$. The mass-weighted metallicity is up to 1.2~dex lower in the outer halo than in the inner halo, whereas this difference is maximally 0.5 and 0.3~dex for the volume- and flux-weighted metallicities, respectively. Because the spread in metallicity is much larger at $z=3$ than at $z=0.25$, the flux-weighted metallicities trace the mass-weighted 84th percentile at $z=3$ as opposed to significantly higher metallicities at $z=0.25$.

\section{Effect of  AGN feedback} \label{sec:AGN}

We repeated the surface brightness profile analysis for simulations including AGN feedback, because this feedback has been shown to be important for the halo gas around massive galaxies \citep[e.g.][]{McCarthy2010, Bertone2010a, Bertone2010b, VoortSchaye2012}. Because the effect of AGN feedback is largest at low redshift and for soft X-ray emission, we only show those results (Figure~\ref{fig:AGNprofileXrayz0p125}), but describe the small differences for UV emission as well.

The surface brightness profiles for soft X-ray lines are shown in Figure~\ref{fig:AGNprofileXrayz0p125} as the dashed curves. The solid curves show the result for the simulations without AGN feedback and are identical to those shown in Figure~\ref{fig:profileXrayz0p125}. 
AGN feedback reduces the emission from $10^{12-13}$~M$_\odot$ haloes by more than an order of magnitude for N~\textsc{vii}, O~\textsc{viii}, and Ne~\textsc{x} in the halo centre. The reduction is smaller at larger radii, but still 0.5~dex at 0.3$R_\mathrm{vir}$. For group-sized haloes, the emission in the central region is also lower by 1~dex for Ne~\textsc{x}, 0.6~dex for N~\textsc{vii} and O~\textsc{viii}, 0.3~dex for C~\textsc{vi}, whereas the O~\textsc{vii} surface brightness in the centre remains the same. At larger radii, the reduction is small, about $0.2-0.3$~dex at 0.5$R_\mathrm{vir}$. In the outer region of cluster haloes, there is no difference when we include AGN. However, in the most inner region, $R<0.1R_\mathrm{vir}$, the surface brightness of all the soft X-ray lines increases by 0.1--0.7~dex when AGN feedback is included.

The predictions for the maximum radius out to which the soft X-ray emission is detectable are not affected much by AGN feedback, except for the lowest halo masses. \textit{ORIGIN} and \textit{IXO/ATHENA} can detect O~\textsc{viii} out to 0.15$R_\mathrm{vir}$, 0.6$R_\mathrm{vir}$, and 0.7$R_\mathrm{vir}$ for haloes with $M_\mathrm{halo}=10^{12-13}$, $10^{13-14}$, and $10^{14-15}$~M$_\odot$, respectively. C~\textsc{vi} and O~\textsc{vii} have surface brightnesses above 0.1~photon~s$^{-1}$~cm$^{-2}$~sr$^{-1}$ within $0.15-0.3R_\mathrm{vir}$. N~\textsc{vii} and Ne~\textsc{x} drop below the detection threshold for $M_\mathrm{halo}=10^{12-13}$.

The analysis for UV and H$\alpha$ emission at $z=0.25$ was repeated for a simulation including AGN feedback, but with lower resolution. We can therefore only compare the two highest mass bins. In the centres of haloes ($R<0.1R_\mathrm{vir}$) with $M_\mathrm{halo}=10^{12-13}$~M$_\odot$ the surface brightness is reduced by 0.5--1~dex. The reduction is largest for C\,\textsc{iv} and O\,\textsc{vi} emission. At larger radii, the reduction is only about 0.2~dex. Surface brightnesses are not affected by AGN feedback for $M_\mathrm{halo}=10^{13-14}$~M$_\odot$ at $R>0.03R_\mathrm{vir}$ and are only slightly reduced at smaller radii.

AGN feedback affects the rest-frame UV surface brightness profiles at $z=3$ the most for haloes with $M_\mathrm{halo}=10^{11-12}$~M$_\odot$, where the reduction in the centre  ($R<0.1R_\mathrm{vir}$) is 0.3~dex for C\,\textsc{iii}, Si\,\text{iii}, and Si\,\textsc{iv} and up to 0.8~dex for C\,\textsc{iv} and O\,\textsc{vi}. At larger radii, the maximum reduction is 0.2~dex. The implemented AGN feedback may be too strong for haloes of this mass for the relatively high resolution of the 25~$h^{-1}$Mpc simulation used at $z=3$, so the actual surface brightnesses could be closer to the ones predicted by the simulation without AGN feedback, as shown in Figure~\ref{fig:profileUVz0p25}. For haloes with $M_\mathrm{halo}=10^{10-11}$~M$_\odot$, the surface brightness profiles are lower (by 0.2--0.5~dex) with AGN feedback, but for haloes with $M_\mathrm{halo}=10^{12-13}$~M$_\odot$, the surface brightness profiles of C\,\textsc{iii}, Si\,\textsc{iii}, and Si\,\textsc{iv} are actually higher with AGN feedback (by 0.2--0.4~dex).

The mass-, volume-, and flux-weighted physical properties we find in the simulations with AGN feedback (not shown) are similar to those in the simulations without AGN feedback. The differences are largest for X-ray emission. The weighted temperatures change by less than 0.1~dex. AGN feedback reduces the median flux-weighted overdensities by 0.1--0.4~dex for most halo masses. Mass- and volume-weighted metallicities increase by 0.1--0.4~dex, but the flux-weighted metallicities increase less when AGN feedback is included.

\section{Summary and discussion} \label{sec:concl}

A large fraction of the gas in galaxy haloes, groups, and clusters has temperatures $T=10^{4.5-7}$~K \citep[e.g.][]{VoortSchaye2012}. At these temperatures, cooling is dominated by metal-line emission for metallicities $Z\gtrsim0.1$~$Z_\odot$ \citep[e.g.][]{Wiersma2009a}. Observing these lines therefore constitutes an excellent route to studying the diffuse, warm-hot gas around galaxies, both at low and at high redshift. Additionally, metal-line emission provides us with clues as to how the circumgalactic gas was enriched and thus about the feedback process responsible for this enrichment.

We used cosmological, hydrodynamical simulations from the OWLS project to quantify the surface brightness profiles of diffuse, circumgalactic gas, for both soft X-ray and rest-frame UV metal lines. This was done by stacking surface brightness maps centred on galaxies in the simulations. We only included emission from gas with densities $n_{\rm H} < 0.1~{\rm cm}^{-3}$, so our predictions should be considered to be lower limits to the actual surface brightness predicted by the models.
The lines we considered are C~\textsc{vi} (0.367 keV), N~\textsc{vii} (0.500 keV), O\textsc{vii} (0.561 keV), O~\textsc{viii} (0.654 keV), and Ne~\textsc{x} (1.021 keV) in the soft X-ray band and C~\textsc{iii} (977 \AA), C~\textsc{iv} (1548 \AA), Si~\textsc{iii} (1207 \AA), Si~\textsc{iv} (1294 \AA), and O~\textsc{vi} (1032 \AA) in the rest-frame UV band. We discussed their detectability with current and future instruments and computed the flux-weighted physical properties of the gas. We also quantified the effect of AGN feedback on our results. 

Proposed X-ray missions with detection limits of $10^{-1}$~photon~s$^{-1}$\,cm$^{-2}$\,sr$^{-1}$ can detect metal-line emission in galaxy haloes, groups, and clusters at $z=0.125$. O\,\textsc{viii} emission is the strongest and can be detected out to 80 per cent of the virial radius of groups ($M_\mathrm{halo}=10^{13-14}$~M$_\odot$) and clusters ($M_\mathrm{halo}=10^{14-15}$~M$_\odot$) and out to 0.4$R_\mathrm{vir}$ for $M_\mathrm{halo}=10^{12-13}$~M$_\odot$. C\,\textsc{vi}, N\,\textsc{vii}, O\,\textsc{vii}, and Ne~\textsc{x} can be detected out to smaller radii, $0.1-0.5R_\mathrm{vir}$. 

Assuming a detection limit of $10^{-20}$~erg~s$^{-1}$\,cm$^{-2}$\,arcsec$^{-2}$ for UV metal lines at $z=0.25$, C~\textsc{iii} can be detected out to 0.3$R_\mathrm{vir}$ for $10^{11-13}$~M$_\odot$ haloes and out to 0.5$R_\mathrm{vir}$ for $10^{13-14}$~M$_\odot$ haloes. It is the strongest UV metal line, but because it has a wavelength blueward of Ly$\alpha$, it may be absorbed by intervening hydrogen clouds. Fortunately, we estimated that the attenuation is small, on average (18 per cent for C~\textsc{iii} emission at $z=3$, see Section~\ref{sec:UV}). C\,\textsc{iv}, O\,\textsc{vi}, Si\,\textsc{iii}, and Si\,\textsc{iv} can be detected out to $0.1-0.2$~$R_\mathrm{vir}$ for halo masses above $10^{11}$~M$_\odot$ (the lowest mass we considered at low redshift), and Si\,\textsc{iii} can even be seen out to the same radius as C~\textsc{iii} for $10^{13-14}$~M$_\odot$ haloes.

Assuming a detection threshold of $10^{-20}$~erg~s$^{-1}$\,cm$^{-2}$\,arcsec$^{-2}$, upcoming optical instruments should be able to detect the rest-frame UV metal lines C\,\textsc{iv}, O\,\textsc{vi}, Si\,\textsc{iii}, and Si\,\textsc{iv} out to $0.1R_\mathrm{vir}$ at $z=3$ for haloes more massive than $10^{11}$~M$_\odot$. C~\textsc{iii} is the strongest emission line and can be observed out to twice that distance. 

The same optical instruments that can detect rest-frame UV metal lines at high redshift, can also observe H$\alpha$ at low redshift. H$\alpha$ is a good probe of the cold ($10^4$~K) halo gas. With a detection threshold of $10^{-20}$~erg~s$^{-1}$\,cm$^{-2}$\,arcsec$^{-2}$, H$\alpha$ at $z=0.25$ may be detected out to 0.2, 0.3, and 0.6~$R_\mathrm{vir}$ for $M_\mathrm{halo}=10^{11-12}$, $10^{12-13}$, and $10^{13-14}$~M$_\odot$, respectively. 

The flux-weighted mean temperatures of the gas probed by the soft X-ray lines increase with halo mass, but less steeply than the mass- and volume-weighted mean temperatures. The emission is biased towards high density gas, especially in the inner halo and for cluster-sized haloes. The volume- and mass-weighted mean metallicities are similar, but always a factor of 3--10 lower than the flux-weighted mean metallicities. The differences between the flux-weighted properties of different metal lines are small for soft X-ray lines, although for clusters O\,\textsc{vi} probes significantly lower temperatures and higher densities than the other metal lines that we consider.

The flux-weighted temperatures of the gas from which the rest-frame UV emission originates is very close to the temperatures at which the corresponding emissivity curves peak, and they are insensitive to halo mass. The temperature of the gas probed is similar 
($T\sim 10^{4.5}-10^5$~K) for all lines, except for O~\textsc{vi}, whose emissivity peaks at $T\sim 10^{5.5}$~K. At low redshift, all the metal lines are biased towards high densities. This behaviour is strongest for the silicon lines and weakest for O~\textsc{vi}. At high redshift, the bias is smaller. In the outer parts of haloes with  $M_\mathrm{halo}<10^{12}$~M$_\odot$, C~\textsc{iv} and O~\textsc{vi} have flux-weighted mean densities below the mass-weighted mean density, but still above the volume-weighted average. The volume- and mass-weighted metallicities are always a factor of 3--30 lower than the flux-weighted values. 

AGN feedback introduces significant uncertainties in the predicted surface brightness profiles, especially at small radii. With AGN feedback, the emission decreases for low-mass haloes, but increases for some of the highest-mass haloes. In most cases, the radius out to which we expect to observe the metal-line emission is, however, not affected much. 

In this work we have not tried to mimic observational errors  and selections effects and we have not accounted for some potentially important physical processes, such as absorption by intervening clouds and emission from the multiphase interstellar medium. Hence, although our results can help build intuition about the level and interpretation of metal-line emission from diffuse gas in haloes, our predictions for the detectability of specific metal lines must be considered highly approximate. Future work could try to improve the realism of the predictions for specific instruments by creating virtual observations. 

Future work could also investigate the usefulness of more sophisticated statistical analyses. For example, stacking randomly oriented haloes, as we have done here, will not show filamentary emission, because the filaments will be oriented randomly. However, before stacking the galaxies, one can first rotate them towards their nearest neighbour of similar mass. This should enhance the emission of the gas in one direction. The signal at large radii is hard to detect, so the brightest emission lines should be chosen for this. We carried out some preliminary tests of this strategy using our simulations, but found that the enhancement is only about 0.2~dex (not shown). It may, however, be worthwhile to investigate other ways to rotate the galaxies before stacking.

Besides stacking all images, it is possible to stack only pixels with a detection in the brightest emission line and then look for emission from other lines. Because the emission is highest in the densest regions, the emission from different lines will be correlated. This will also lead to an enhancement of the signal. Pixels with a detection in Ly$\alpha$ could, for example, be well-suited for detecting C~\textsc{iii}, C~\textsc{iv}, Si~\textsc{iii}, and Si~\textsc{iv} as these metal-lines probe relatively cold ($T\sim 10^4 - 10^{5}$~K) gas. We intent to investigate such a strategy in future work.

\section*{Acknowledgements}

We would like to thank Serena Bertone for help with the construction of the emissivity tables and Jelle Kaastra and all the members of the OWLS team for valuable discussions. We would also like to thank the anonymous referee for his or her useful comments. The simulations presented here were run on Stella, the LOFAR BlueGene/L system in Groningen, on the Cosmology Machine at the Institute for Computational Cosmology in Durham as part of the Virgo Consortium research programme, and on Darwin in Cambridge. This work was sponsored by the National Computing Facilities Foundation (NCF) for the use of supercomputer facilities, with financial support from the Netherlands Organization for Scientific Research (NWO), also through a VIDI grant. The research leading to these results has also received funding from the European Research Council under the European Union's Seventh Framework Programme (FP7/2007-2013) / ERC Grant agreement 278594-GasAroundGalaxies and from the Marie Curie Training Network CosmoComp (PITN-GA-2009-238356).

\bibliographystyle{mn2e}
\bibliography{emission}

\bsp

\appendix
\section{Resolution tests}

We tested the dependence of our results on pixel size, bin size of the surface brightness profile, and on the thickness of the region for which the emission is added. We found that the pixel size and bin size are unimportant for our results. If the thickness of the region is increased by up to an order of magnitude, the surface brightness outside the virial radius is increased by at most 0.3~dex. These surface brightnesses are below the unresolved X-ray background and therefore not observable with proposed instruments.

\begin{figure*}
\center
\includegraphics[scale=.8]{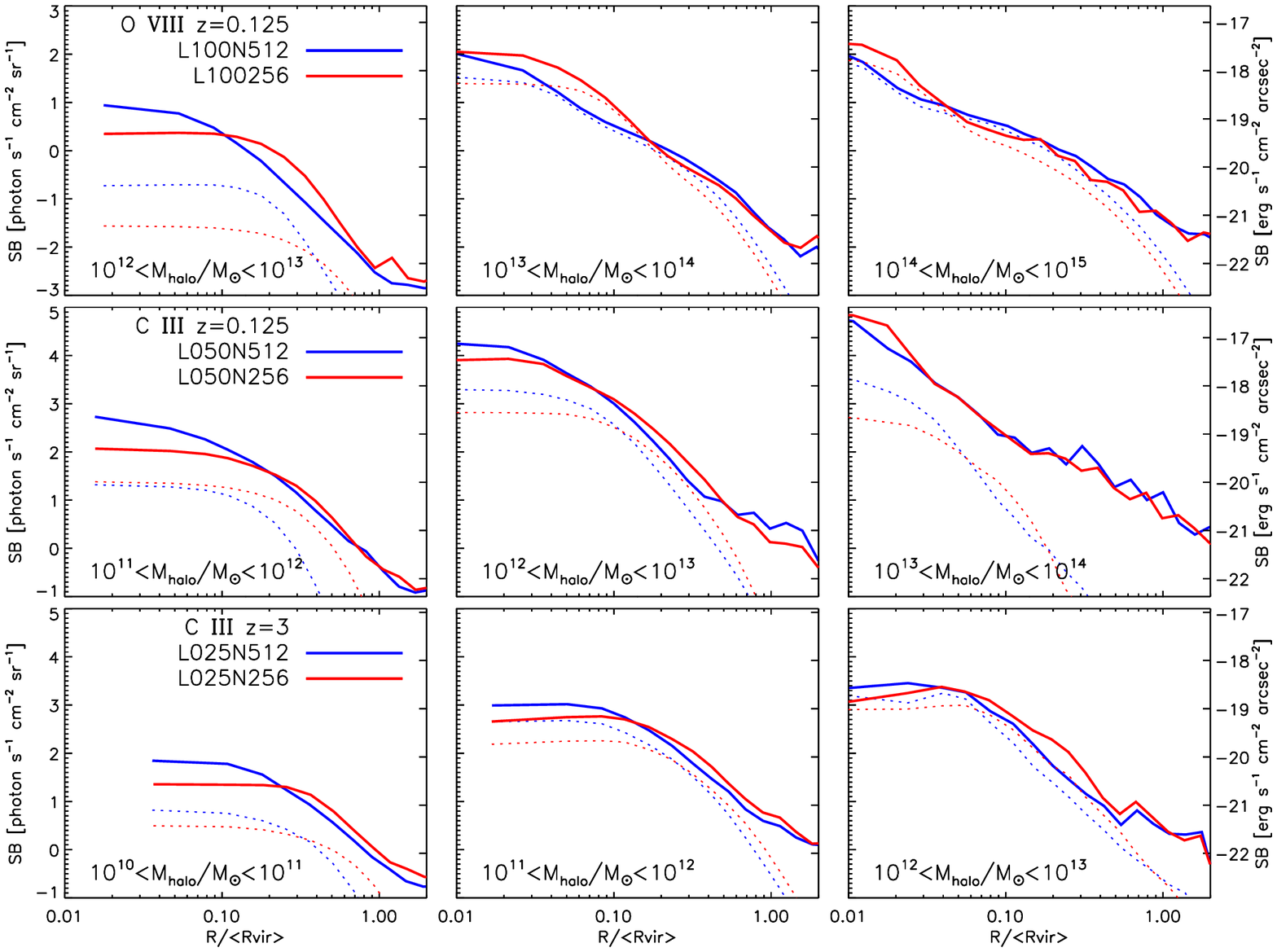}
\caption {\label{fig:profileres} Mean (solid curves) and median (dotted curves) O~\textsc{viii} surface brightness at $z=0.125$ (top row), C~\textsc{iii} surface brightness at $z=0.25$ (middle row), and C~\textsc{iii} surface brightness at $z=3$ (bottom row) as a function of radius for the same haloes as in Figures~\ref{fig:profileXrayz0p125}, \ref{fig:profileUVz0p25}, and \ref{fig:profileUVz3p0}, respectively. Note that different rows show different halo mass bins. The blue curves show the profiles for high-resolution simulations, as also used throughout this paper. The red curves show the profiles for simulations with eight (two) times lower mass (spatial) resolution.}
\end{figure*}

Our results do not depend on the box size of the simulation, but there is a small dependence on resolution, which is shown in Figure~\ref{fig:profileres}. Blue curves show the original surface brightness profiles for the brightest line in Figures~\ref{fig:profileXrayz0p125}, \ref{fig:profileUVz0p25}, and \ref{fig:profileUVz3p0}. Red curves show the results for simulations with eight (two) times lower mass (spatial) resolution. Differences for the other X-ray and UV lines are very similar and thus not shown. The largest discrepancies between the low- and high-resolution simulation are found at small radii.

\label{lastpage}

\end{document}